\documentclass[journal=jacsat,manuscript=article]{achemso}

\usepackage[version=3]{mhchem} % Formula subscripts using \ce{}
\usepackage{subcaption}
\usepackage{booktabs}
\usepackage{siunitx}
\usepackage{textgreek}
\usepackage{bm}

\author{Peter Pak}
\affiliation{
  Department of Mechanical Engineering, Carnegie Mellon University, Pittsburgh,
  PA, USA
}

\author{Victor Alvarado}
\affiliation{
  Department of Mechanical Engineering, Carnegie Mellon University, Pittsburgh,
  PA, USA
}
\author{Amir Barati Farimani}
\email{barati@cmu.edu}
\affiliation{
  Department of Mechanical Engineering, Carnegie Mellon University, Pittsburgh,
  PA, USA
}
\alsoaffiliation{
  Machine Learning Department, Carnegie Mellon University, Pittsburgh, PA, USA
}

\title[]{AI Agentic Selective Laser Sintering Process Optimization}

\abbreviations{IR,NMR,UV}
\keywords{American Chemical Society, \LaTeX}

\SectionNumbersOn
\begin{document}

\begin{abstract}
Agentic systems enable the intelligent automation of complex workflows, specific
to additive manufacturing this is applicable for complex tasks such as process
parameter optimization for mechanical properties. This work investigates the AI
enabled agentic process optimization within Selective Laser Sintering (SLS) to
iteratively improve the tensile and flexural properties of 3 different materials
on the Inova Mk1. These materials include PA12 GF, PA11 Onyx, and PA12 Blend
(volume mixture of 25\% PA12 GF and 75\% PA12 White) and with using knowledge
from previous builds and minimal guidance from the user, the agentic system was
able to optimize process parameters over a small number of iterations to achieve
comparable TDS specified mechanical properties. This work showcases the ability
for an agentic system to continually learn from updated data, enabling the
intelligent automation of complex tasks such as process parameter optimization
for selective laser sintering.
\end{abstract}


\section{Introduction}

Additive Manufacturing (AM) enables rapid iteration and development cycles,
alleviating the need for rigid tooling, specialized fixtures, and traditional
manufacturing requirements \cite{beaman_additive_2020, zhao_laser_2023,
kumar_selective_2003}. Within additive manufacturing each process presents its
own unique set of advantages such as relatively minimal postprocessing for Fused
Deposition Modeling (FDM) \cite{ngo_additive_2018}, high precision and
functional materials of Laser Powder Bed Fusion (LPBF) \cite{zhao_laser_2023}
and E-Beam \cite{galati_literature_2018}, and large form factor of Wire Arc
Additive Manufacturing (WAAM) \cite{li_comprehensive_2022}. These advantages are
effectively realized under ideal process parameters, however, nuances between
environments and machines prohibit simple utilization of recommended settings
\cite{brown_interlaboratory_2016}.

Agentic systems enable the intelligent automation of complex tasks such as the
development of functional assemblies \cite{pak_rocketsmith_2026,
barkley_cadsmith_2026}, drug discovery \cite{ock_large_2025}, computational
materials science \cite{chaudhari_modular_2026}, software development
\cite{han_tdflow_2026}, and alloy design \cite{pak_agentic_2026,
ghafarollahi_automating_2025, ghafarollahi_rapid_2025}. These agentic systems
are particularly capable of search and optimization by enabling Large Language
Models (LLMs) to operate within dynamic environments through the use of tools
and data resources \cite{pak_agentic_2026, pak_rocketsmith_2026}. Tools can
include software platforms, Application Programming Interfaces (APIs), or system
level commands which provide the LLM with greater capability to interact with
its environment. With these capabilities, agentic systems not only enable tool
orchestration and reasoning but further augment the user's ability to execute
upon complex and multi-faceted challenges.

Additive Manufacturing (AM) is a suitable domain where agentic systems are quite
applicable \cite{pak_agentic_2026, jadhav_llm-3d_2025}. Existing work for build
planning, defect mitigation, and process optimization rely on multi-physics
simulations \cite{hemmasian_surrogate_2023, ogoke_thermal_2021}, machine
learning \cite{ogoke_deep_2024, bostan_accurate_2025, pak_thermopore_2024}, and
specialized domain knowledge. Powder based processes such as Selective Laser
Sintering (SLS) and Laser Powder Bed Fusion (LPBF) require precision controlled
environments where parameters such as chamber temperature, laser power, scanning
velocity, and numerous other factors can have an affect on part quality
\cite{niu_selective_2000, kumar_selective_2003, zhao_laser_2023}. With the
proper process parameters, LPBF and SLS achieve micron level precision and are
uniquely capable of producing complex geometric features that are otherwise
impossible through other manufacturing processes \cite{zhao_laser_2023,
kumar_selective_2003}. However, the search for optimal process parameters is a
tedious undertaking since factors such as varying hardware components like
optics introduce an additional source of variability. Calibration of a single
set of process parameters requires multiple cycles of fabrication, testing, and
analysis; presenting a gap where agentic systems can execute in a more efficient
manner \cite{ahmed_process_2022, beuth_process_2013}.

\begin{figure}[htbp]
    \centering
    \includegraphics[width=\textwidth]{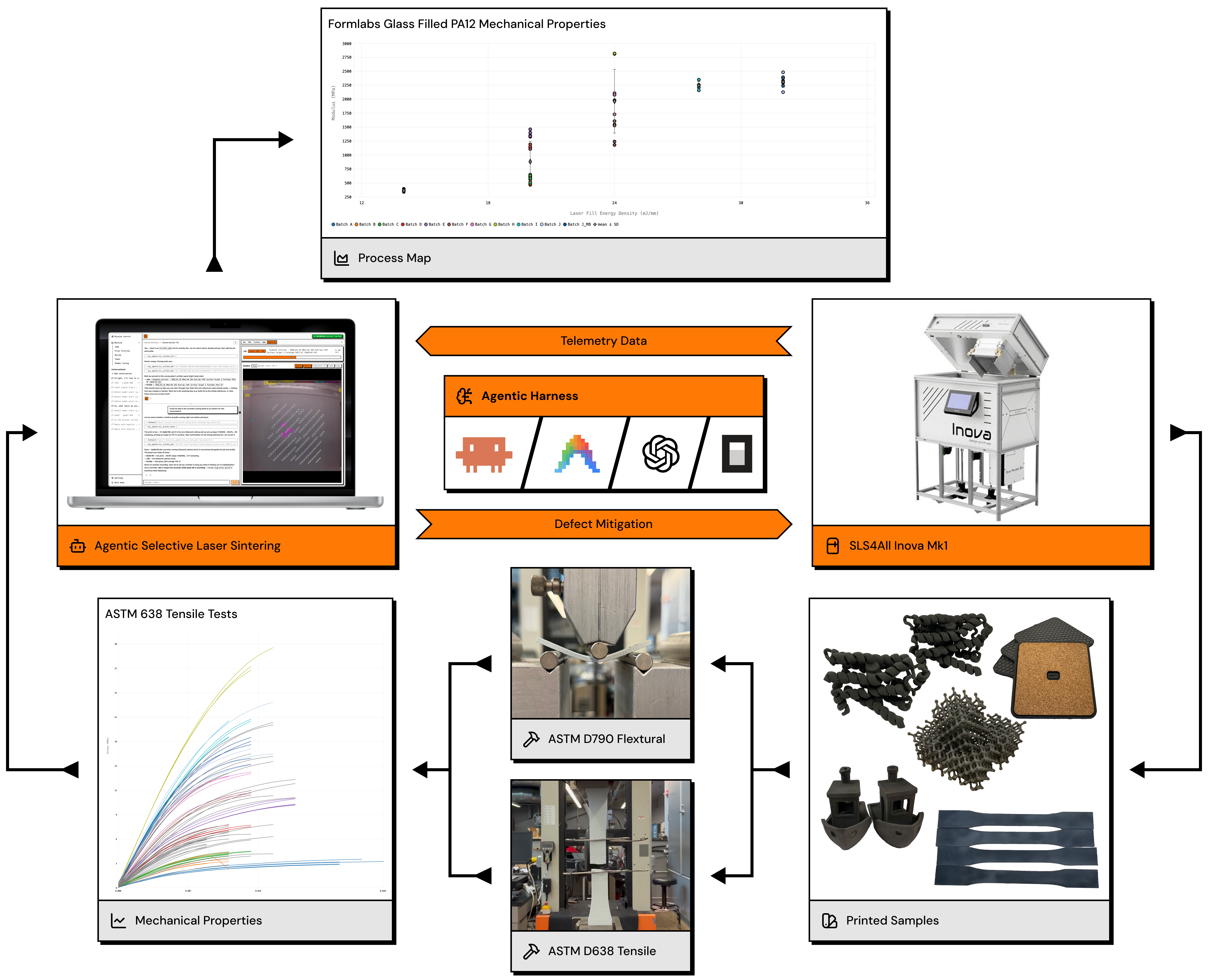}
    \caption{
    Agentic system for selective laser sintering capable of process parameter
    optimization and continual learning through evaluation of printed ASTM
    samples.
    }
    \label{fig:agentic_sls_main_figure}
\end{figure}

This work introduces an agentic system for the optimization of process
parameters in selective laser sintering evaluated to the ASTM standard on a
variety of materials. The system is capable of operating with a
human-in-the-loop approach where utilizing ASTM test data from previous builds,
firmware level tool calls, and reasoning enabled by Large Language Models (LLM),
an optimal set of process parameters which result in desired mechanical
properties can be efficiently determined. Samples are fabricated using the
SLS4All Inova Mk1, an open source SLS machine built from kit, allowing for the
collection of telmetry data in the form of optical image, surface temperature,
and positional data. The collected data is compiled into a dataset utilized for
further training of ancillary models and granting additional reasoning context
enabling a continual learning agentic system (Figure
\ref{fig:agentic_sls_main_figure}).

\section{Related Work}

LLM-3D Print by Jadhav et al. \cite{jadhav_llm-3d_2025} is an early work that
explores the application of a large language model to the task of process
monitoring within the additive manufacturing process of Fused Deposition
Modeling (FDM). This work developed a LangChain based agentic system which
through a top down optical camera image data regarding the immediately printed
layer using a 1.0 mm nozzle \cite{jadhav_llm-3d_2025}. The system utilized these
top down images to detect layer defects and address with corrective parameter
changes in subsequent layers with various agents orchestrated by GPT-4o. These
agents are responsible for actions such as gathering information, planning, and
executing solutions \cite{jadhav_llm-3d_2025}. With this system authors were
able to show that an LLM is effective in detecting and compensating major defect
factors such as stringing, oozing, and layer separation on par of that of a
human. Furthermore, compression tests on samples constructed with and without
the multi-agent system achieved a max peak load 5x greater than the same sample
constructed without the system \cite{jadhav_llm-3d_2025}. This work provides the
groundwork for an agentic system in additive manufacturing capable of addressing
build layer defects through in-situ monitoring realized through superior
mechanical properties in the final part.

Agentic additive manufacturing alloy evaluation focuses on the use of LLM
enabled tool calling for the task of developing a lack of fusion process map
within laser powder bed fusion for a proposed alloy composition
\cite{pak_agentic_2026}. This work enables an agentic system to utilize software
platforms such as Thermo-Calc to generate material properties for various alloy
compositions which are then utilized in a Rosenthal based thermal model to
determine melt pool dimensions utilized in lack of fusion process map
calculations \cite{pak_agentic_2026}. For known materials such as Inconel 718
and Stainless Steel 316L, the system exhibited good alignment of the lack of
fusion process regimes of various layer heights with the values found in the
literature \cite{pak_agentic_2026}. For unknown alloy composition, the predicted
lack of fusion process regimes displayed a general trend to match the general
lack of fusion defect regime expected with compositions such as 99\% Iron \&
1\%, Copper, and Al-Si-10Mg \cite{pak_agentic_2026}. This agentic system
showcases the large language model's ability to reason through complex user
queries for and ultimately generate concrete, data based process maps.

Specific to materials science, Chaudhari et al. \cite{chaudhari_modular_2026}
developed a multi-agent framework named MatSciAgent composed of 4 different
agents capable of materials retrieval, continuum simulation, crystal structure
generation, and molecular dynamics simulation. The authors utilize GPT-3.5-Turbo
as the core orchestrator of each agent and demonstrates the capability of each
with a number of case studies. AtomAgents \cite{ghafarollahi_automating_2025}
investigates the use of a multi-agent system equipped with physics based
simulation tools to overcome the limitations the LLM's training data. The
authors explore this within 4 different case studies regarding materials
property calculation, dislocation analysis, multi-scale mechanical problem
solving, and hypothesis generation and validation. Ghafarollahi et al.
\cite{ghafarollahi_rapid_2025} extends upon this and explores the use of a Graph
Neural Network (GNN) enabled agentic system for the discovery of new potential
alloy candidates within the NbMoTa family of alloys. In this search, candidates
are optimized under the properties of Peierls barrier and screw dislocation
energy which their GNN model predicts allowing the system to autonomously
navigate through the design space of alloys considering both atomic-scale
material properties and macro-scale mechanical properties
\cite{ghafarollahi_rapid_2025}.

\section{Methodology}
\label{sec:methodology}

\subsection{Experimental Platform}
\subsubsection{SLS4All Inova Mk1}
\begin{figure}[htbp]
    \centering
    
    \begin{subfigure}[b]{0.48\textwidth}
        \centering
        \includegraphics[width=\textwidth]{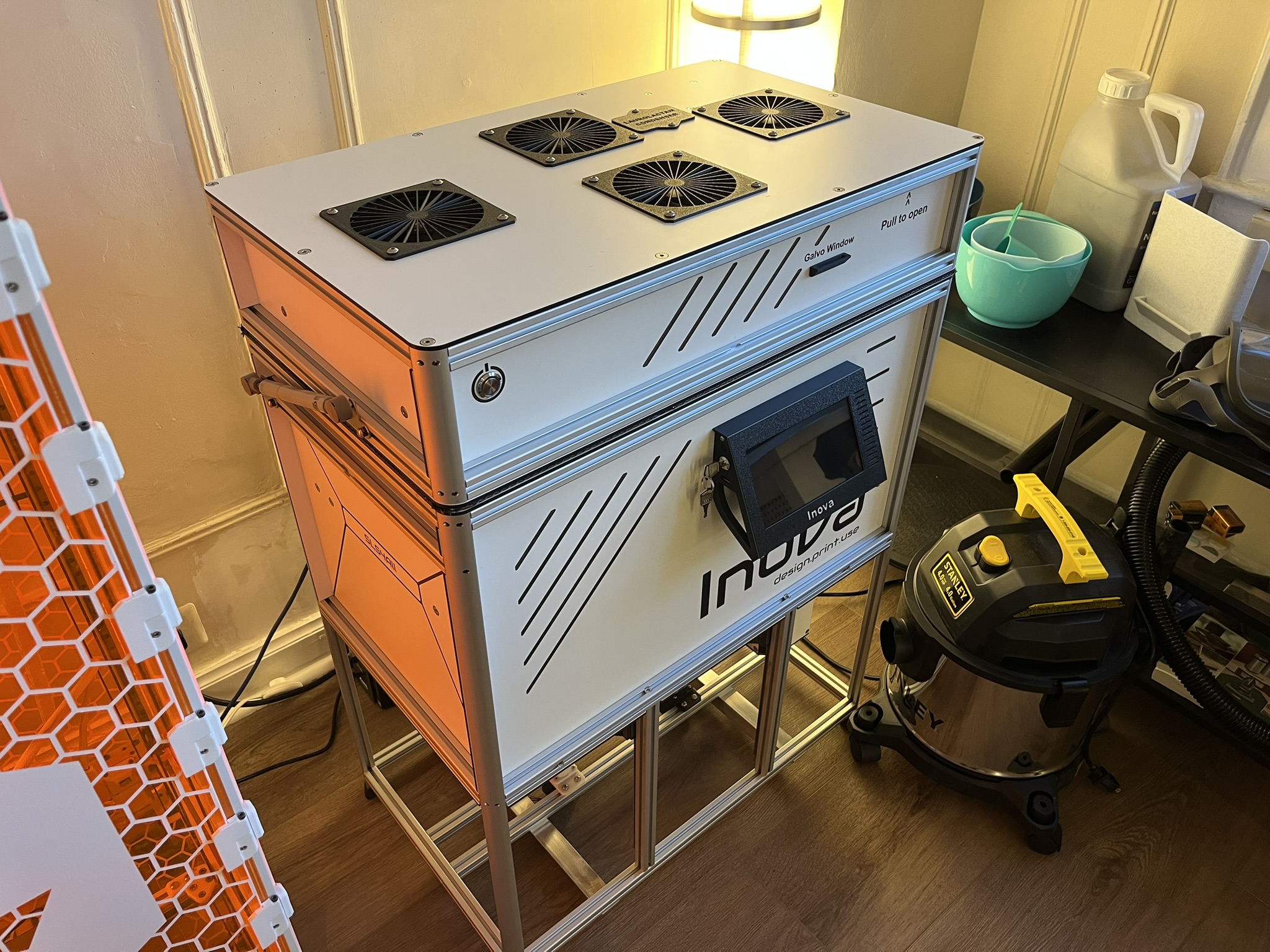}
        \caption{SLS4All Inova Mk1}
        \label{fig:inova_mk1}
    \end{subfigure}
    \hfill
    \begin{subfigure}[b]{0.48\textwidth}
        \centering
        \includegraphics[width=\textwidth]{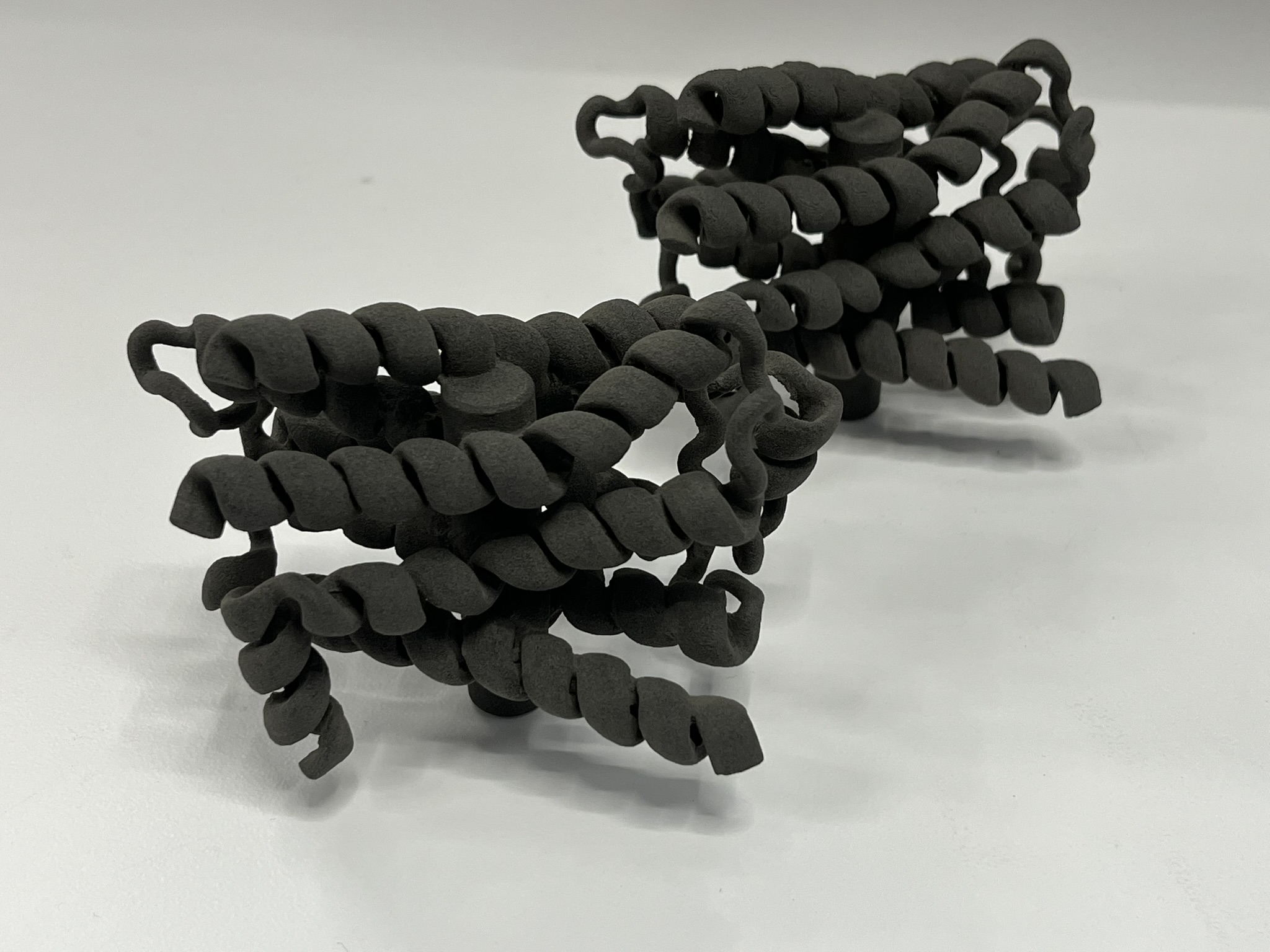}
        \caption{Printed \textbeta 2 Protein Model}
        \label{fig:inova_print_proteins}
    \end{subfigure}
    
    \caption{
    Inova Mk1 (a) assembled from kit prints complex geometries, such as the
    \textbeta2 protein molecule (b), only capable through the SLS process.
    }
    \label{fig:inova_mk1_and_print}
\end{figure}

The selective laser sintering process will be performed using the Inova Mk1
(Fig. \ref{fig:inova_mk1}) for a range of materials such as PA12, PA12 GF, and
other experimental powder compositions. The Inova Mk1 is an open source, low
cost, selective laser sintering machine developed by SLS4All founders Tomas
Starek and Pavel Dyntera \cite{starek_sls4all_2020}. This machine was purchased
as kit from SLS4All and assembled over the course of several months producing
successful prints (Fig. \ref{fig:inova_print_proteins}) using Formlab's PA12 GF
\cite{formlabs_nylon_2026}. The Inova Mk1 utilizes a 450 nm blue diode laser
capable of delivering 10 watts of power \cite{starek_sls4all_2020}. The machine
is capable of an effective build volume of 150 mm x 150 mm x 185 mm and utilizes
an array of 4 halogen lamps for surface heating control and a 5 heating elements
for build chamber temperature control \cite{starek_sls4all_2020}. Surface
temperature monitoring and control is achieved with a \textit{ThermoCam
Waveshare MLX90640} capable of producing a 32 x 24 pixel thermal image for
temperatures ranging from 0 \textdegree C to 300 \textdegree C. The optical
camera utilizes an Omnivision OV5647 sensor capable of streaming a 1920 x 1080
pixel image at 30 frames per second. Average scan speed is around 1,650 mm/s at
5 watts (250 \textmu m spot size) and 2,800 mm/s at 10 watts (350 \textmu m spot
size) \cite{starek_sls4all_2020}. The firmware controlling the hardware
components such as the galvometers, laser, stepper motors, sensors, and heating
elements runs off a combination of open-source software programs including
\texttt{Klipper} \cite{kevinoconnor_klipper3dklipper_2026},
\texttt{SLS4All.Compact} \cite{dyntera_sls4allsls4allcompact_2025}.

\subsubsection{Platform Limitations}
The Inova Mk1 system presents several software and hardware limitations.
Although the software is primarily open-source, modules disclosing specific
feature implementations such as part slicing and tool path generation are
considered proprietary and only provided in a compiled state. In addition, there
is little documentation regarding the machine's exposed Application Programming
Interface (API) as the dashboard web interface is streamed from the host. The
maintainers do offer an option for utilizing custom firmware plugins and this is
the primary method in which the agentic system communicates with the Inova Mk1.
With regard to the installed sensors, relative to thermal and optical imaging
equipment used in other in-situ process monitoring works
\cite{myers_high-resolution_2023, myers_two-color_2023, pak_thermopore_2024,
bostan_accurate_2025, ogoke_deep_2024}, the resolution and frame rate that these
sensors provide is comparatively coarse. This adds a potential constraint to the
quality of real-time information that can be utilized by the agentic system when
testing process parameters and executing builds. The manufacturer also suggest
material restrictions to primarily polymer based powders with a melting
temperature of 200 \textdegree C \cite{starek_sls4all_2020}.

\subsection{Agentic System}
\begin{figure}[htbp]
    \centering
    \includegraphics[width=\textwidth]{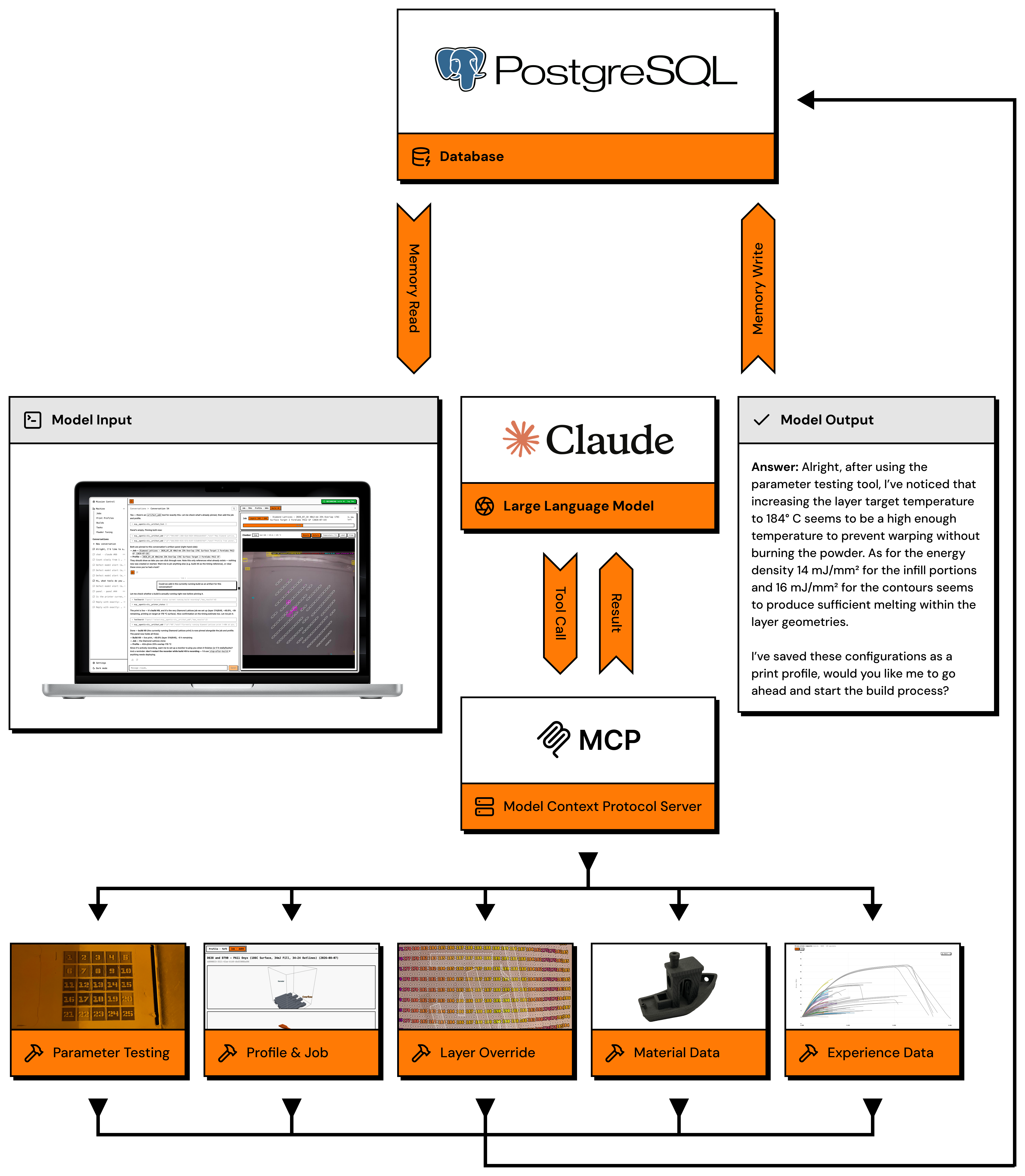}
    \caption{
        Agentic Selective Laser Sintering (SLS) process begins with user input
        for determining process parameters for a proposed material sent to the
        Large Language Model (Claude Fable). Query utilizes data from previous
        conversations stored in relational database (PostgreSQL) and selection
        of tools via Model Context Protocol (MCP) to investigate query. Model
        generates a response, saved for reference in future tasks enabling a
        continual learning environment.
    }
    \label{fig:agentic_sls_flow}
\end{figure}

The agentic system (Figure \ref{fig:agentic_sls_flow}) enables the intelligent
automation of parameter selection and process monitoring through the guidance of
large language model. Integration with a relational database (i.e. PostgreSQL)
enables the use of dynamic memory such that previous system outputs and tool
call results can be recalled for future prompts, providing additional context
for the large language model to use during reasoning. Lastly, the system is able
to interact with its environment through agentic tool calls following the Model
Context Protocol (MCP). This includes general knowledge tools such as those
specific to material properties, previous builds, print profile configuration,
and ASTM testing data. In addition, the system is also capable of executing
machine firmware level tool calls which allow for the creation and adjustment of
process parameters for the next ASTM build and runtime overrides for surface
temperature, recoating, and energy.

\subsubsection{Agent Harness}
The agent harness is the framework which implements the functionality overhead
for enabling the LLM to communicate and execute tool calls, create and maintain
user conversations, and other various system level tasks. For this agentic
system a collection of model harnesses including Claude Code, Antigravity CLI,
OpenAI Codex, and OpenCode are implemented with the Claude Fable model used to
execute process parameter optimization. Through the standard of the Model
Context Protocol, all of the tools and implemented functionality can be utilized
interchangeably with each of the agent harnesses enabling a modular system which
can easily replace the system's current LLM with one that better suits the need
of the user.

\begin{figure}[htbp]
    \centering
    \includegraphics[width=0.75\textwidth]{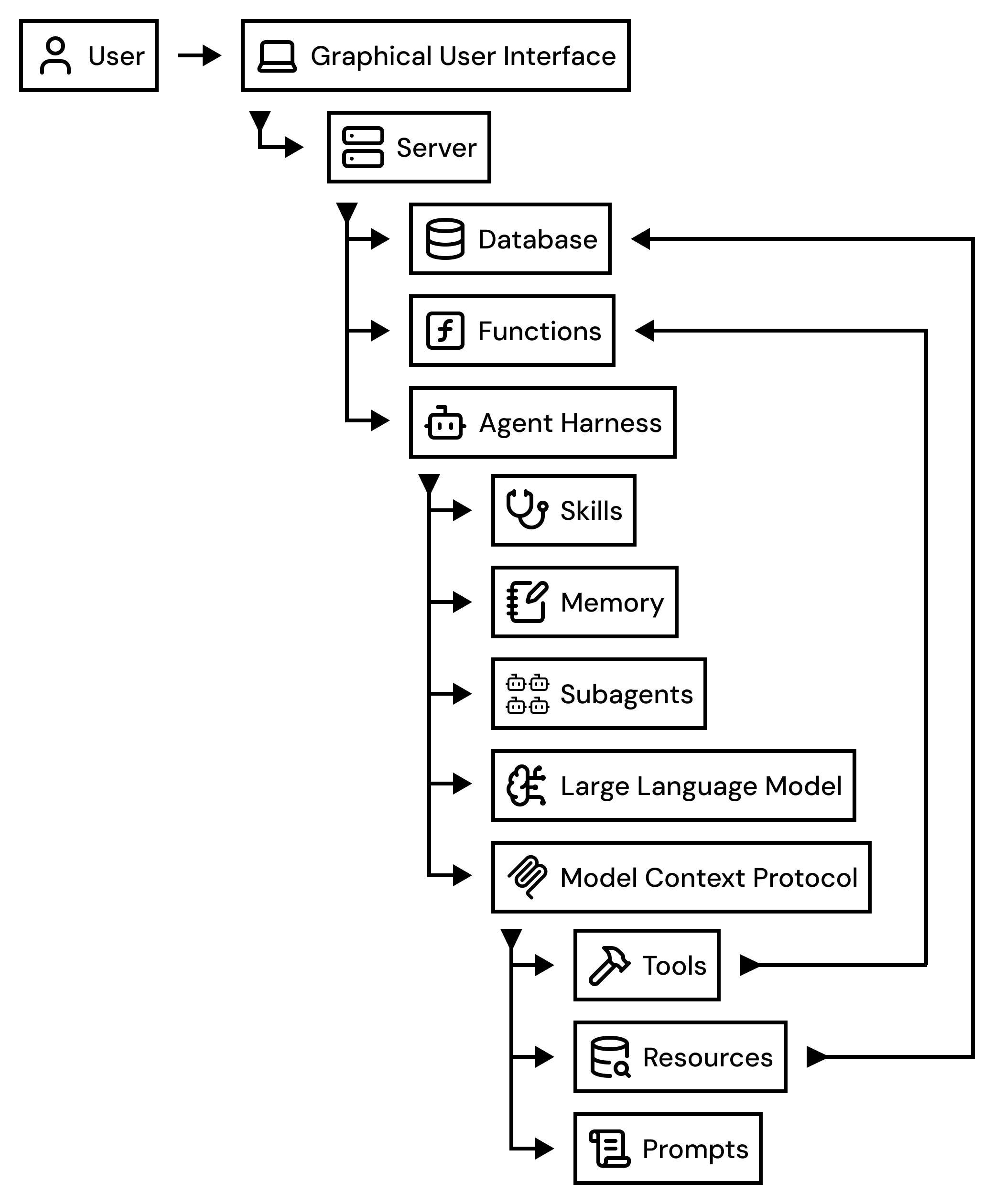}
    \caption{
        The agent harness provides the large language model with the
        functionality to utilize tools, data, and memory to better interact with
        its environment. A common architecture is outlined above where tools and
        resources allow for access into server hosted functions and database
        access respectively.
    }
    \label{fig:model_harness}
\end{figure}

\subsubsection{MCP Enabled Tool Calling}

Tool calling is orchestrated by the LLM and enabled through the model harness
via the Model Context Protocol. For this agentic system the tools can be
organized into several group based on their functionality, those being: material
information, initial testing, print profile and build history, recorded data,
and runtime overrides.

\begin{figure}[htbp]
    \centering
    
    \begin{subfigure}[b]{0.56\textwidth}
        \centering
        \includegraphics[width=\textwidth]{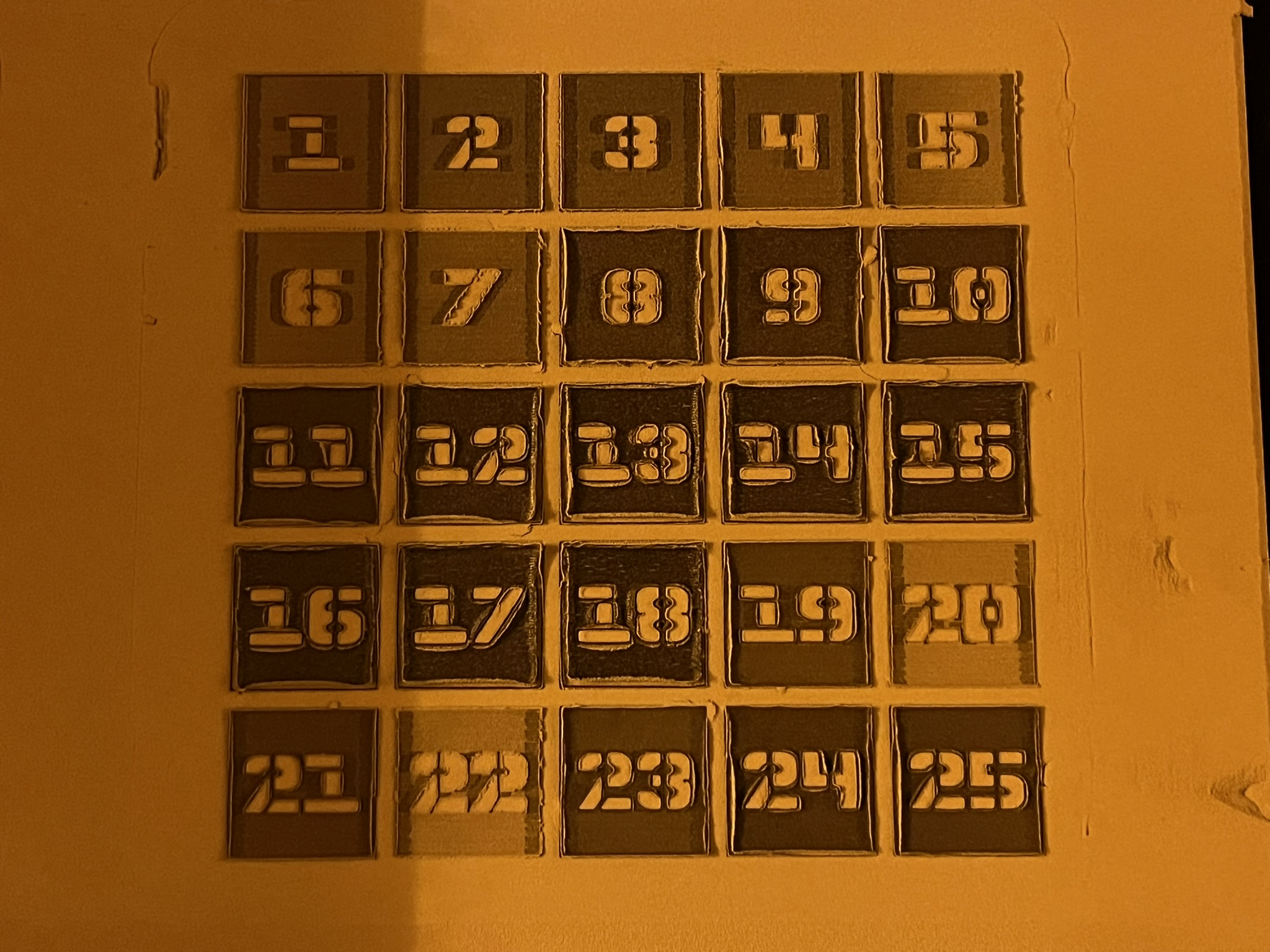}
        \caption{Printed Parameter Testing Patches}
        \label{fig:inova_parameter_testing}
    \end{subfigure}
    \hfill
    \begin{subfigure}[b]{0.4025\textwidth}
        \centering
        \includegraphics[width=\textwidth]{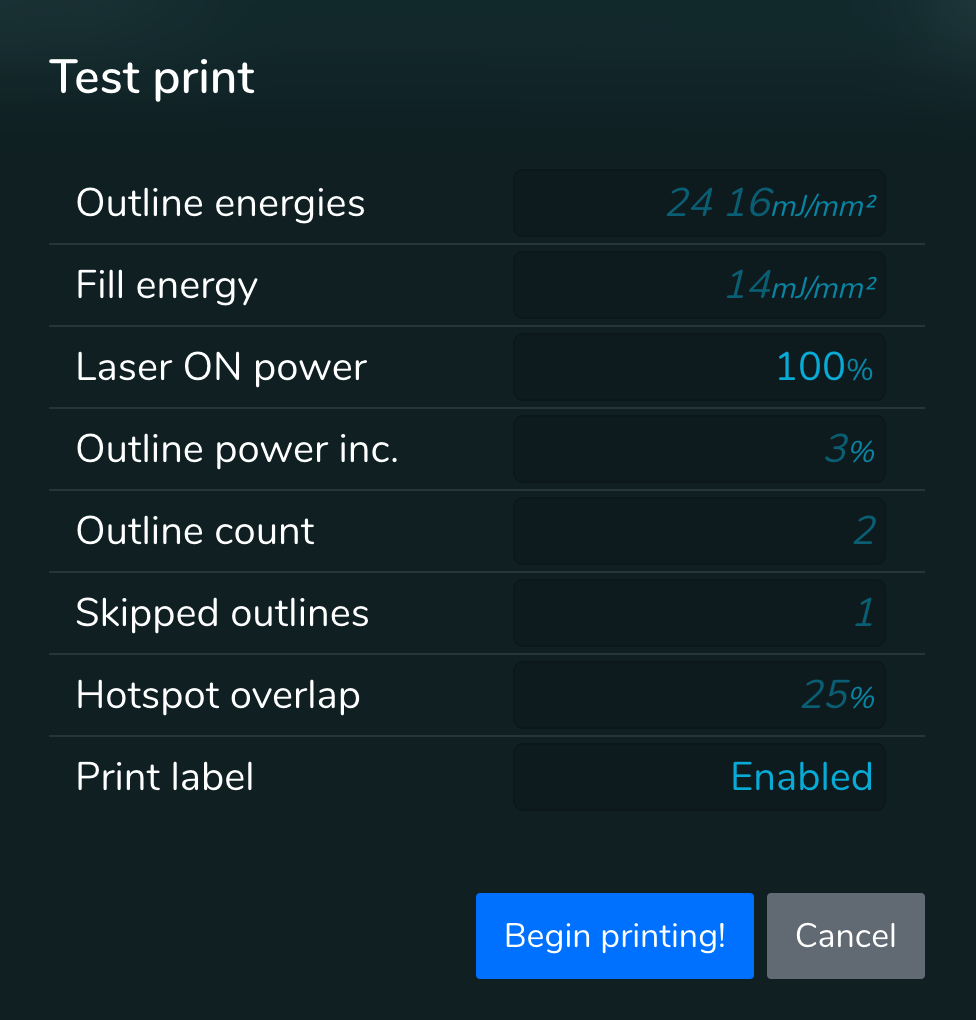}
        \caption{Patch Configuration}
        \label{fig:inova_parameter_testing_config}
    \end{subfigure}
    
    \caption{
        5 x 5 grid of patches (a) printed for testing PA12 GF powder, each with
        their own parameter configuration (b).
    }
    \label{fig:inova_mk1_parameter_testing}
\end{figure}

Within the realm of material information, tools here allow for inference on
specified materials and their approximate properties. For example in the case of
PA12 GF, tools repsective to materials would query upon manufacturer published
technical data sheets, machine level process parameters, and other available
data to provide the necessary information required to experiment with this
material. The parameter testing tool is a feature specific to the Inova Mk1 such
that when experimenting with new powders, a 5 x 5 grid of labeled patches
(Figure \ref{fig:inova_parameter_testing}) can be printed on the surface of the
print bed. This allows for the rapid testing of print parameters through patch
specific configurations such as outline and fill energy densities (Figure
\ref{fig:inova_parameter_testing_config}). In addition to executing patch
prints, the tool will utilize the available sensors to extract quality
information regarding each printed patch to determine which patch configuration
to test next.

\noindent For jobs and print profiles, the tools here set the various build
parameters for a specific material within the Inova Mk1 for future use. Notable
parameters include surface and chamber temperatures, energy densities for the
fill and contours, hotspot overlape, desired layer height, and recoater speed.
The runtime overrides tools allow for the LLM to control various aspects of the
build including surface temperature, recoating passes, and applied energy to
address potential issues that may occur duing the build. Lastly tools related to
recorded data allows for the LLM to investigate the effectiveness of previous
process parameter settings such that adjustments can be applied to improve
mechanical performance measured through ASTM testing.

\subsection{Manufacturing Process}

With prescribed print parameters and job templates, each build will be printed
with a consistent powder chamber preheat of 145 \textdegree C, 100 \textmu m
layer height, 100\% recoater speed, and cooling procedure. Print parameters are
maintained and executed by the Inova Mk1 firmware to the best of its ability.
For example array of halogen lamps placed over the print surface to maintain a
consistent even surface temperature (halogens are pulsed to increase and reduce
surface temperature), however, the realized surface temperature may differ.
After each build the print is left to cool down overnight (often 8+ hours)
before the powder cake is removed for post processing. Powder is sifted and
reclaimed during postprocessing and material is kept separated to the best
ability of the author.

\subsection{Sample Evaluation}

\begin{figure}[htbp]
    \centering
    \includegraphics[width=0.75\textwidth]{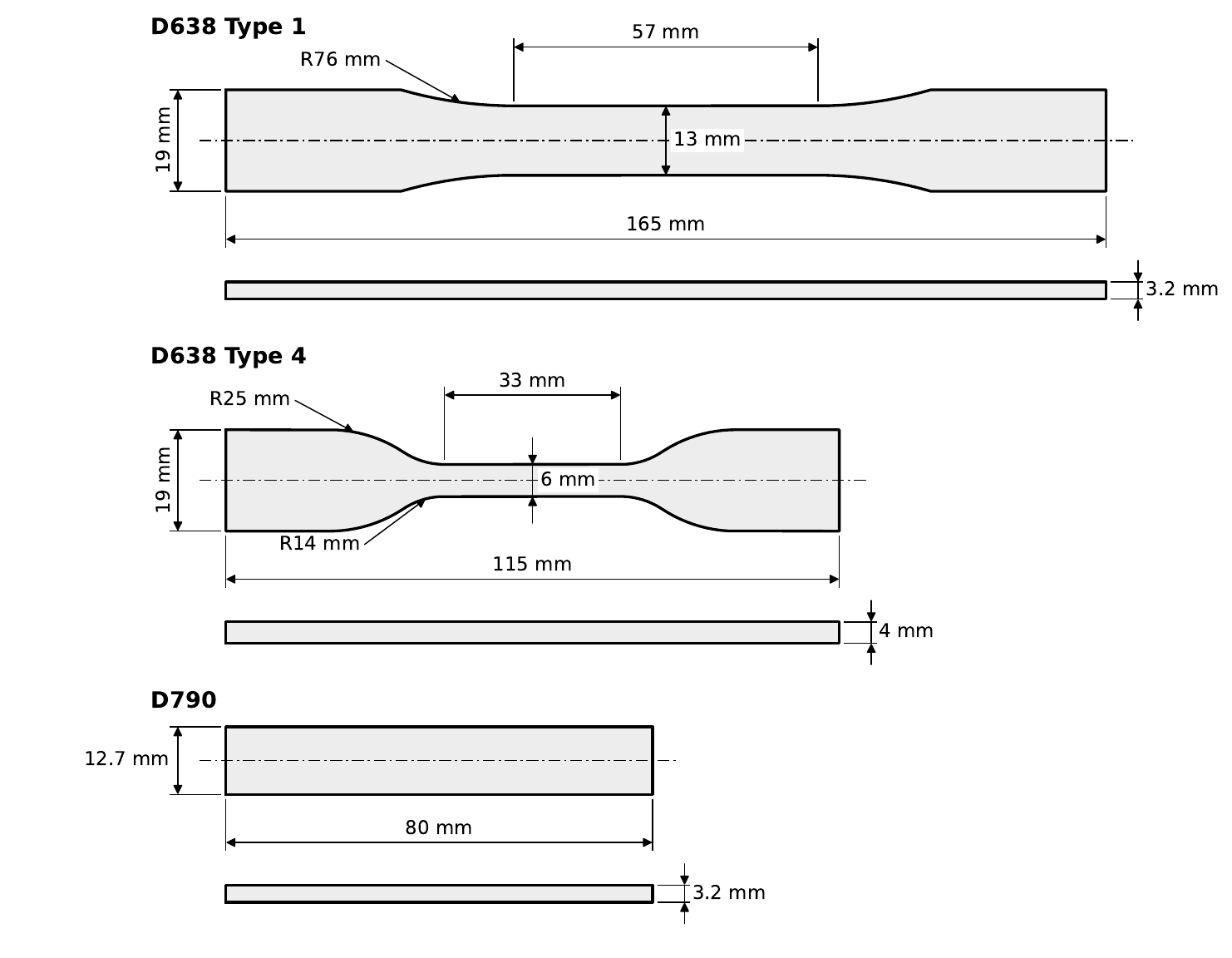}
    \caption{
        ASTM D638 and D790 samples dimesions used in the evaluation of
        mechanical properties.
    }
    \label{fig:agentic_sls_all_samples_drawing}
\end{figure}

Testing samples (Figure \ref{fig:agentic_sls_all_samples_drawing}) were
constructed to the standards outlined by the American Society for Testing and
Materials (ASTM) and compared to measurements published by the manufacturer.
With regard to tensile properties, specifications from the ASTM D638
\cite{d20_committee_test_2022} standard (Type 1 and Type 4) will be used to
construct and test the samples. For bending stress analysis, specifications from
the the ASTM D790 \cite{d20_committee_test_2025} standards will be followed.
Apart from their respective batch-wise process parameters, all samples are
fabricated with a layer height of 100 \textmu m, powder chamber temperature of
145 C, and the major axis of each sample parallel to the print bed to minimize
build height. For each batch, a minimum of 5 samples were printed to adequately
obtain a range of values for benchmarking. Tensile and flexural testing was
conducted with the Instron 4469 Universal Testing System using a 50kN load cell
following the prescribed methodology within the respective standard.

\section{Results}
\subsection{Material 1 (Nylon 12 Glass Filled)}
The control series of tests were conducted using Formlabs Glass Filled Nylon 12
(PA12 GF) powder, manually tuned over the course of 15 different batches. After
the initial assembly of the Inova Mk1, PA12 GF was utilized as the primary
material used to calibrate the various sensors, optical configuration, and
temperature control of the machine. During this period, the basic selective
laser sintering capability was displayed through the fabrication of initial
prints (Figure \ref{fig:inova_print_proteins}), although these parts lacked the
mechanical properties expected from PA12 GF.

At this baseline, a laser fill energy density of 14 mJ/mm with outline of 24
mJ/mm and 16 mJ/mm was used to fabricate the part with a hotspot overlap of 25\%
and a surface bed temperature of 168 C. With these process parameter settings
(Batch A) an initial tensile modulus of around 365 MPa was recorded,
significantly lower than the published 2800 MPa from the manufacturer's TDS. In
addition to the properties listed in the TDS, reference samples were obtained
from manufacturer (Form Now printing service from Formlabs) to physical verify
the listed mechanical properties. These samples were printed on the Formlabs
Fuse series of SLS machines which utilize a 30 W fiber laser in contrast to the
Inova's 10 W diode laser. These reference samples exhibited as tensile modulus
of around 2599 MPa with a spread of 98 MPa and a flexural modulus of 1950 MPa
with a spread of 271 MPa.

Process parameters and build testing layouts were then manually tuned over these
series of batches as to match the mechanical properties of the reference samples
with the Inova Mk1 (Table \ref{tab:agentic-sls-material-1}). Batch J achieves
the highest tensile properties closest to that specified by the manufactured at
a fill energy density of 32 mJ/mm, a hotspot overlap of 50\%, and a surface
temperature of 178 \textdegree C.  Batch H does achieve a higher tensile modulus
of 2815 \textpm 4 MPa and a flexural modulus of 2272 \textpm 396 MPa, however
these properties were not reproducible and attributed to potential measurement
error. This build used a print surface temperature target of 173 C, fill energy
density of 24 mJ/mm (outline energy densities of 28 mJ/mm and 18 mJ/mm), and a
hotspot overlap of 50\%. Through a set of manual tuning explorations, a
collection of baseline data was obtained to further inform the agentic system in
future parameter exploration tasks as seen with material 2 and material 3.

\begin{table}[t]
  \centering
  \begin{tabular}{lccccccc}
    \toprule
    & \multicolumn{3}{c}{Select print parameters} & \multicolumn{4}{c}{Mechanical properties} \\
    \cmidrule(lr){2-4} \cmidrule(lr){5-8}
    Batch & Fill & Overlap & $T_\mathrm{surface}$ & $E_t^{*}$ & UTS$^{*}$ & $E_f^{*}$ & $\sigma_f^{*}$ \\
    & (mJ/mm) & (\%) & (\si{\celsius}) & (MPa) & (MPa) & (MPa) & (MPa) \\
    \midrule
    A   & 14 & 25 & 172 & $397$  & $3.5$  & ---    & ---    \\
    B   & 20 & 25 & 172 & $609$  & $4.4$  & ---    & ---    \\
    C   & 20 & 25 & 172 & $651$  & $4.6$  & $404$  & $11.0$ \\
    D   & 20 & 50 & 173 & $1193$ & $8.0$  & $938$  & $22.4$ \\
    E   & 20 & 50 & 173 & $1402$ & $11.2$ & $1232$ & $26.9$ \\
    F   & 24 & 50 & 173 & $1609$ & $12.1$ & $1560$ & $32.3$ \\
    G   & 24 & 50 & 173 & $1981$ & $14.3$ & $1634$ & $34.0$ \\
    H$^{\ddagger}$ & 24 & 50 & 173 & $2817$ & $29.6$ & $2773$ & $68.5$ \\
    I   & 28 & 50 & 173 & $2349$ & $21.6$ & $1684$ & $43.7$ \\
    J   & 32 & 50 & 178 & $\bm{2482}$ & $\bm{23.0}$ & $1177$ & $24.2$ \\
    K   & 32 & 50 & 173 & $2036$ & $16.7$ & $1233$ & $28.1$ \\
    L   & 40 & 25 & 173 & $1187$ & $7.8$  k $\bm{2130}$ & $\bm{55.4}$ \\
    M   & 40 & 25 & 176 & $1002$ & $6.9$  & $915$  & $20.6$ \\
    N   & 24 & 50 & 173 & $1040$ & $6.7$  & $770$  & $13.6$ \\
    \bottomrule
    \multicolumn{8}{l}{\footnotesize $^{*}$Best-specimen value of tested
      ASTM D638 (\(E_t\), UTS) and D790 (\(E_f\), \(\sigma_f\)) samples.} \\
    \multicolumn{8}{l}{\footnotesize $^{\ddagger}$Not reproducible, removed from consideration.} \\
  \end{tabular}
  \caption{
    Manually tuned print profile settings and respective tensile and flexural
    properties obtained through ASTM tests Formlabs PA12 GF.
  }
  \label{tab:agentic-sls-material-1}
\end{table}

\subsection{Material 2 (Nylon 11 Onyx)}

Nylon 11 Onyx (PA11 Onyx) from Sinterit presents a unique challenge to for the
task of process parameter optimization as the material exhibits ideal mechanical
properties when printed in a nitrogen rich environment, intended for the Lisa
series of printers. From the manufacturer specification, the powder is expected
to exhibit a tensile and strength of 1680 MPa and 55 MPa respectively along with
a flexural modulus and strength of 1290 MPa and 54.2 MPa. With these
specifications, the agentic system presented with the task of searching for
optimal process parameters to set for a build utilizing the material PA11 Onyx.

\begin{table}[t]
  \centering
  \begin{tabular}{lccccccc}
    \toprule
    & \multicolumn{3}{c}{Select print parameters} & \multicolumn{4}{c}{Mechanical properties} \\
    \cmidrule(lr){2-4} \cmidrule(lr){5-8}
    Batch & Fill & Overlap & $T_\mathrm{surface}$ & $E_t^{*}$ & UTS$^{*}$ & $E_f^{*}$ & $\sigma_f^{*}$ \\
    & (mJ/mm) & (\%) & (\si{\celsius}) & (MPa) & (MPa) & (MPa) & (MPa) \\
    \midrule
    O   & 36  & 50  & 186 & $934$  & $15.7$ & ---    & --- \\
    P   & 32  & 33  & 186 & $827$  & $16.6$ & $425$  & $16.6$ \\
    Q   & 32  & 50  & 186 & $1323$ & $22.1$ & $867$  & $31.3$ \\
    R   & 34  & 50  & 188 & $\bm{1571}$ & $\bm{36.3}$ & $\bm{1093}$ & $\bm{42.5}$ \\
    \bottomrule
    \multicolumn{8}{l}{\footnotesize $^{*}$Best-specimen value of tested
      ASTM D638 (\(E_t\), UTS) and D790 (\(E_f\), \(\sigma_f\)) samples.} \\
  \end{tabular}
  \caption{
    Agentically optimized print profile settings and respective tensile
    properties for PA11 Onyx achieves similar mechanical properities outlined in
    TDS from Sinterit.
  }
  \label{tab:agentic-sls-material-2}
\end{table}

\begin{figure}[htbp]
    \centering
    \includegraphics[width=0.75\textwidth]{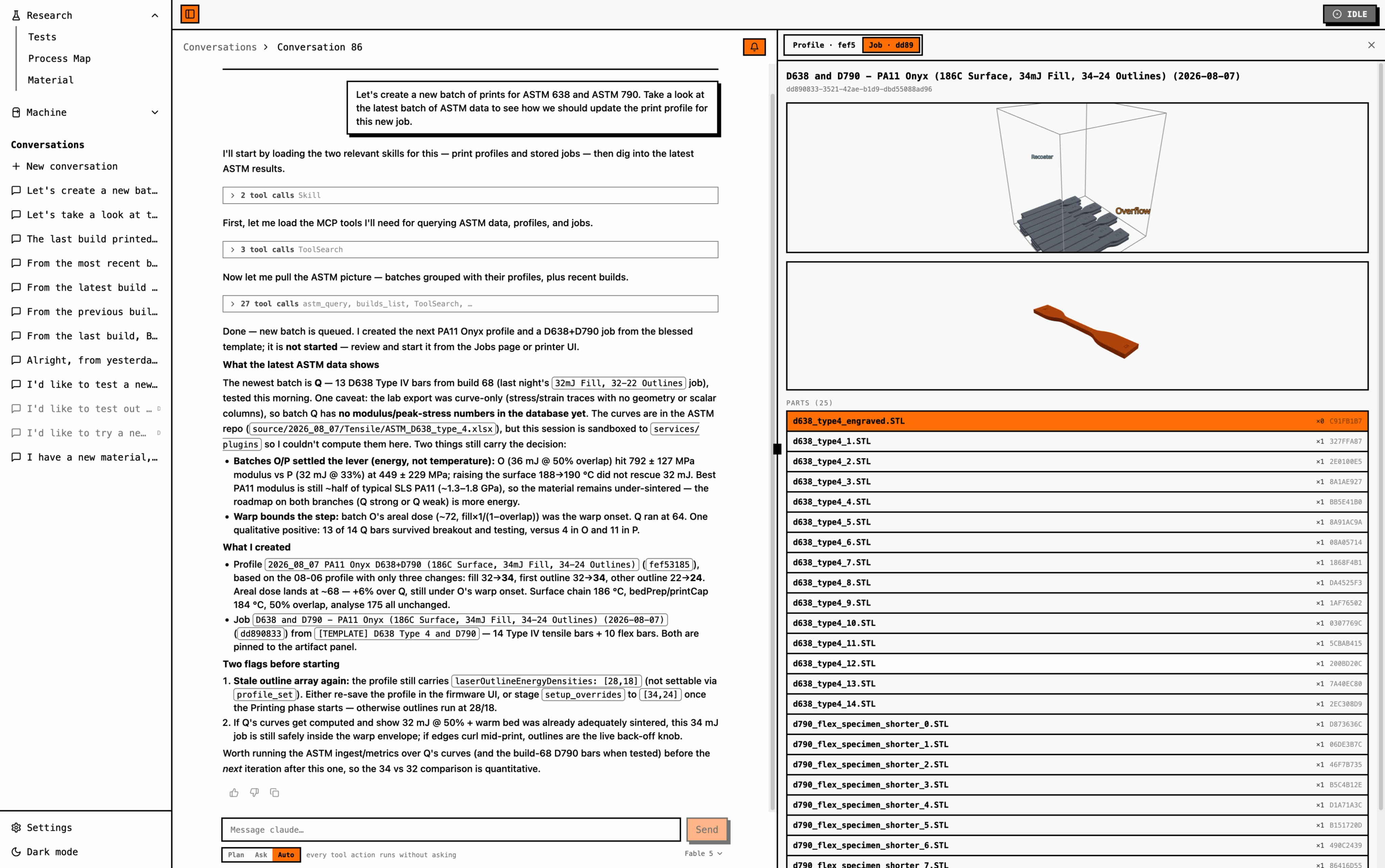}
    \caption{
        GUI showcasing the user flow of build planning to investigate a novel
        set of process parameters along with the reasoning from the agentic
        system.
    }
    \label{fig:agentic_sls_batch_r_gui}
\end{figure}

For each subsequent batch the system is prompted to recall previous builds and
through a series of tool calls the LLM investigates the relationship between
process parameters and the respective mechanical properties of the tested
samples. A Graphical User Interface (GUI) is provided to the user to view the
prescribed process parameters along with the model's reasoning before starting
the build (Figure \ref{fig:agentic_sls_batch_r_gui}).

\begin{figure}[htbp]
    \centering
    
    \begin{subfigure}[b]{0.33\textwidth}
        \centering
        \includegraphics[angle=-90,width=\textwidth]{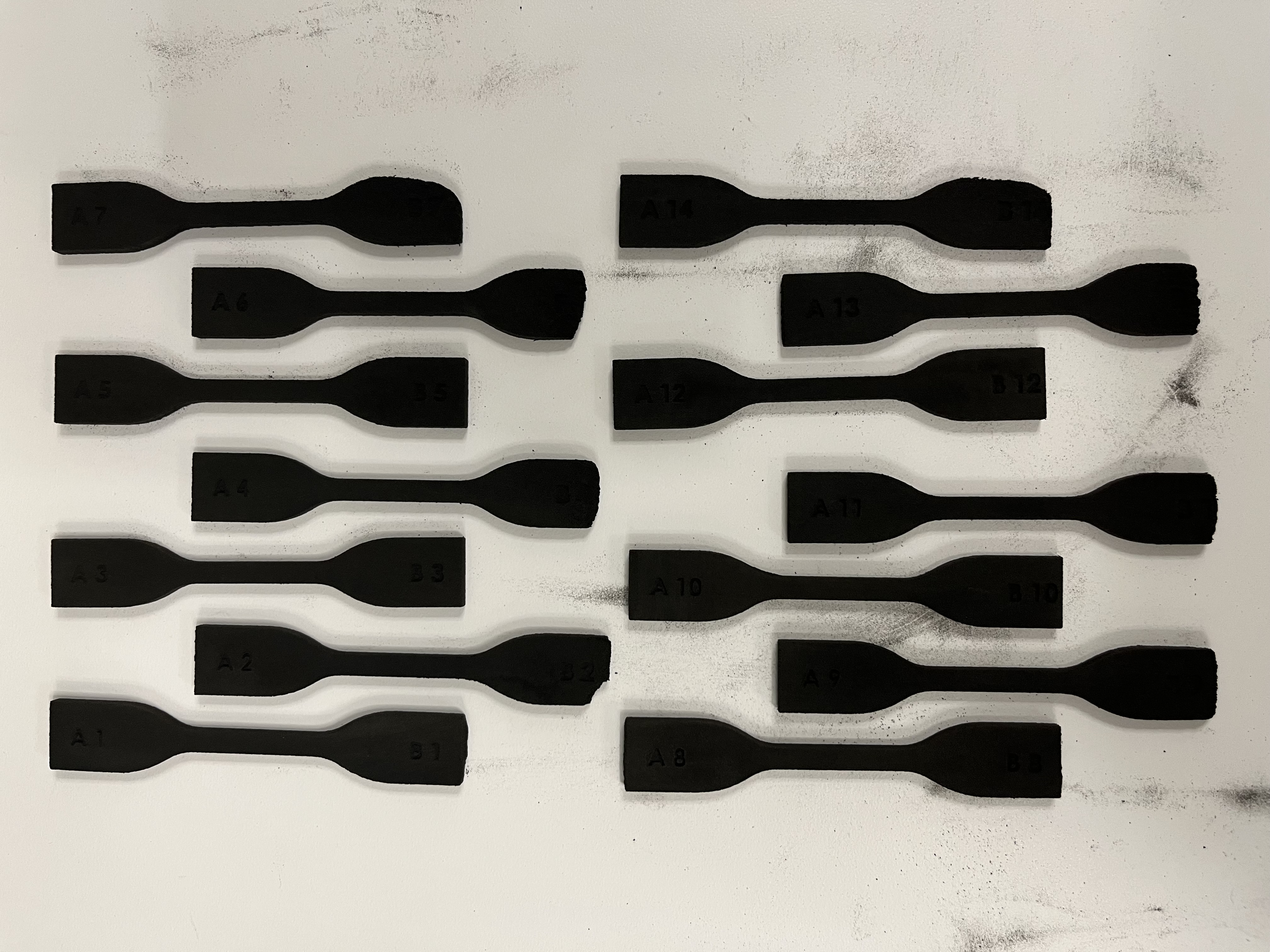}
        \caption{Batch Q}
        \label{fig:agentic_sls_batch_q_tensile_samples}
    \end{subfigure}
    \begin{subfigure}[b]{0.33\textwidth}
        \centering
        \includegraphics[angle=-90,width=\textwidth]{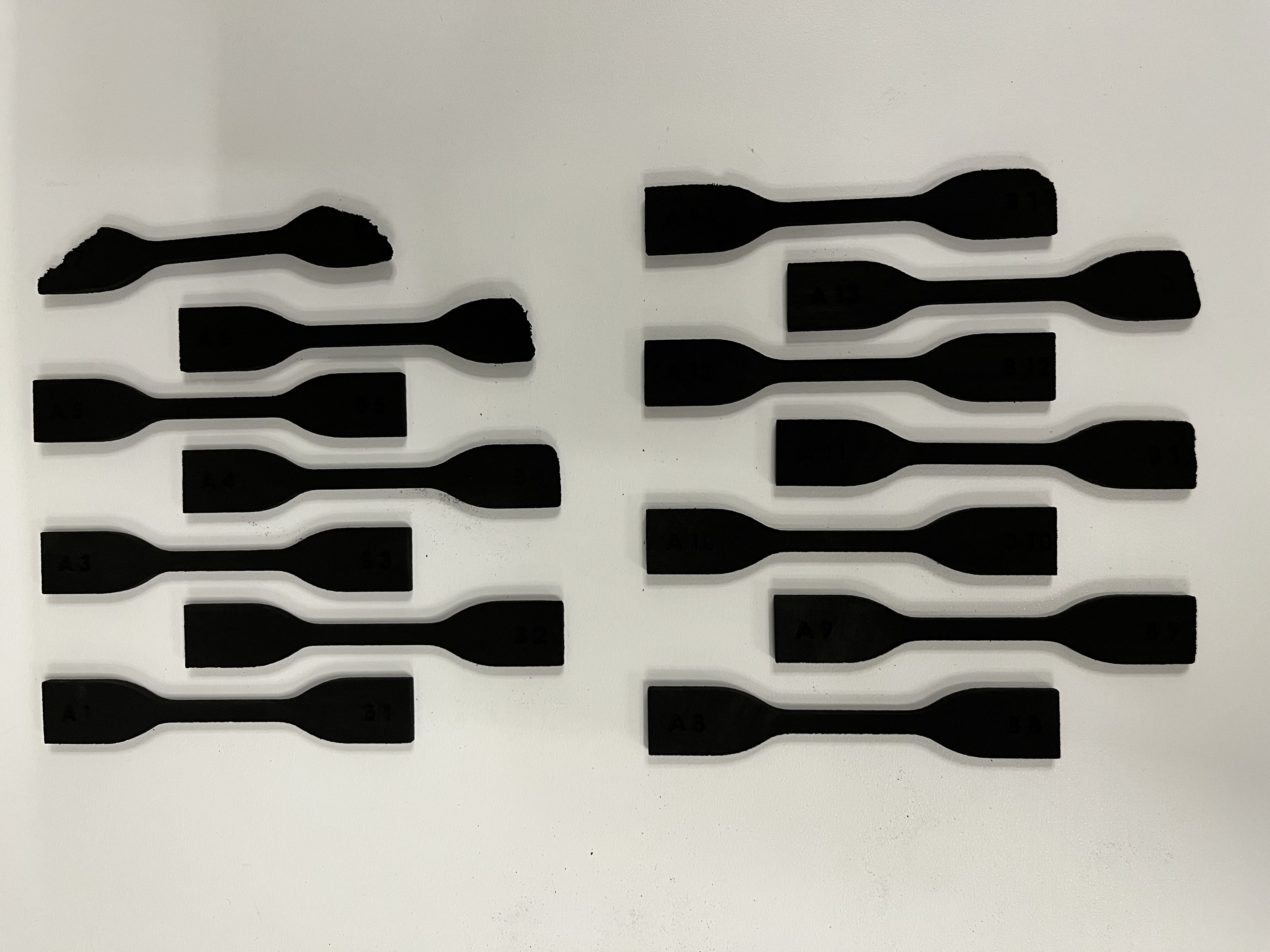}
        \caption{Batch R}
        \label{fig:agentic_sls_batch_r_tensile_samples}
    \end{subfigure}
    
    \caption{
        Tensile samples for Batch Q and R printed with PA11 Onyx. Top row
        includes samples 1 - 7 and bottom row showcases samples 8 - 14.
    }
    \label{fig:agentic_sls_pa11_onyx_tensile_samples}
\end{figure}

\begin{figure}[htbp]
    \centering
    \begin{subfigure}[b]{0.48\textwidth}
        \centering
        \includegraphics[width=\textwidth]{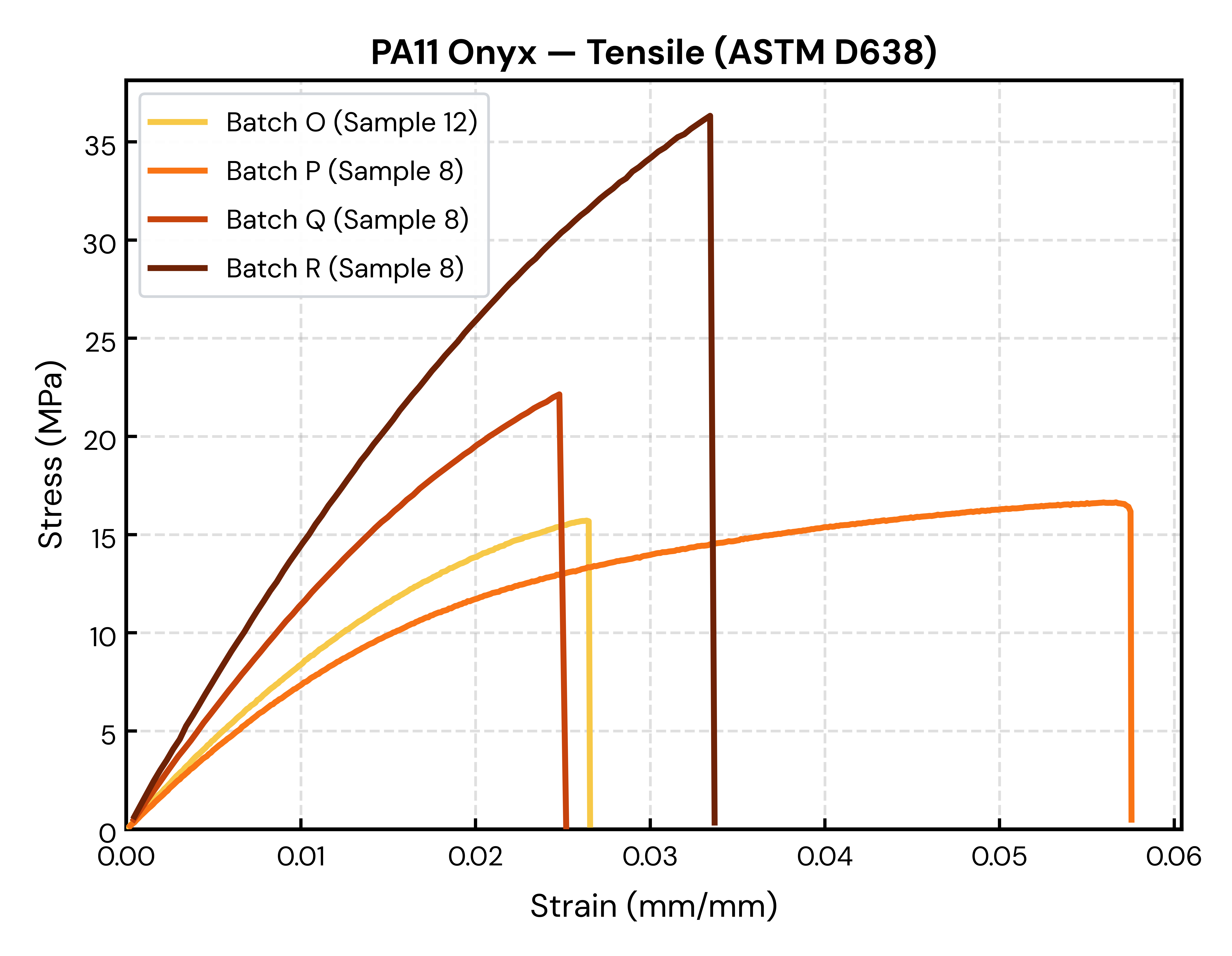}
        \caption{PA11 Onyx Tensile Samples}
        \label{fig:agentic_sls_material_2_tensile}
    \end{subfigure}
    \hfill
    \begin{subfigure}[b]{0.48\textwidth}
        \centering
        \includegraphics[width=\textwidth]{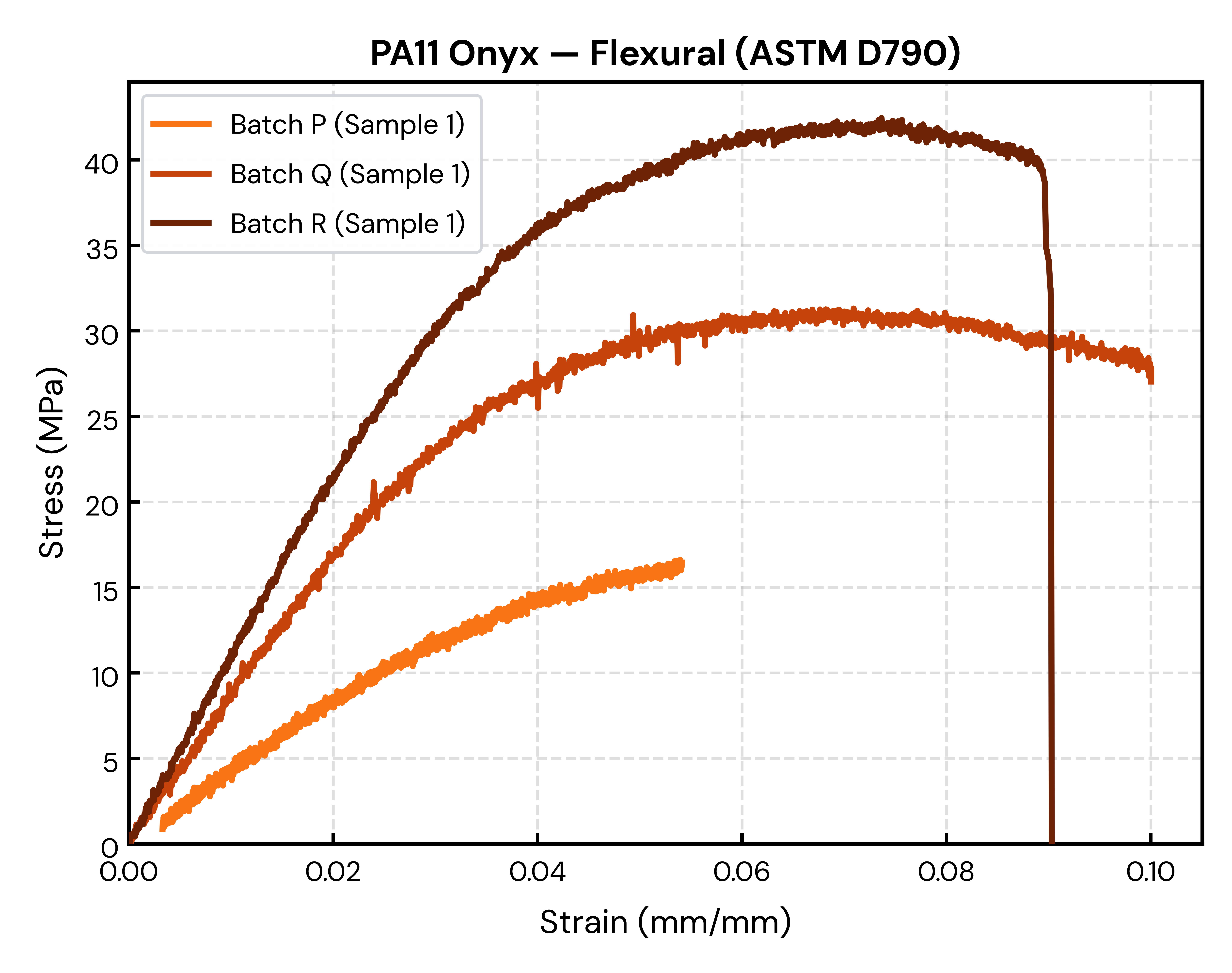}
        \caption{PA11 Onyx Flexural Samples}
        \label{fig:agentic_sls_material_2_flexural}
    \end{subfigure}
    \caption{
        PA11 Onyx tensile and flexural properties show gradual improvement with
        each subsequent batch, best sample from Batch R.
    }
    \label{fig:agentic_sls_material_2_batch_comparisons}
\end{figure}

After initial testing conducted using the parameter testing tool, batches
investigating the process parameters and mechanical characteristics for PA11
Onyx were printed. These include batches O, P, Q, and R which are compiled in
Table \ref{tab:agentic-sls-material-2}. Batch O investigated a fill energy
density of 36 mJ/mm along with outlines energy densities of 36 mJ/mm and 26
mJ/mm with a hotspot overlap of 50\%. This print encountered significant warping
during the build process and only 3 samples were successfully obtained and
tested for this build with the highest tensile modulus of 934 MPa. The next
batch of samples (P) were printed with a hotspot overlap of 33 \% and a fill
energy density of 32 mJ/mm and outline energy densities of 32 mJ/mm and 22
mJ/mm, a significant downstep from the previous sample. For the highest achieved
values, this resulted in a lower tensile modulus of 827 MPa and flexural modulus
of 425 MPa. Batch Q (Figure \ref{fig:agentic_sls_batch_q_tensile_samples}) kept
the same energy density configuration but revisited the hotspot overlap of 50\%
and achieved successful results with the highest achieved tensile modulus of
1323 MPa. Batch R (Figure \ref{fig:agentic_sls_batch_r_tensile_samples}) further
increases the fill energy density to 34 mJ/mm and increase the surface
temperature from 186 C to 188 C and achieved a tensile modulus of 1571 MPa. With
the results from each batch, the agentic system shows its capability to learn
from the previous batch of results and improve the mechanical properties with
subsequent prints (Figure \ref{fig:agentic_sls_material_2_batch_comparisons}).

\subsection{Material 3 (Nylon 12 Blend)}
Material 3 is a mixture of Formlabs Glass Filled Nylon 12 (PA12 GF) and Formlabs
White Nylon 12 (PA12 White), approximately 25\% and 75\% respectively by volume.
PA12 White exhibits lower absorptivity than its PA12 GF counterpart further
observed in failed parameter testing runs where sufficient sintering was not
observed while utilizing the energy density settings of the Inova's 450 nm blue
laser. Previous trials with PA12 GF proved successful sintering with elevated
temperatures and energy densities indicating this powder exhibits an
absorptivity suitable for the Inova's optical configuration. The Nylon 12 Blend
(PA12 Blend) explores the utilization of PA12 GF as a dopant to increase the
base absorptivity of PA12 White in order to achieve sintering and optimize to
ideal mechanical properties.

\begin{table}[t]
  \centering
  \begin{tabular}{lccccccc}
    \toprule
    & \multicolumn{3}{c}{Select print parameters} & \multicolumn{4}{c}{Mechanical properties} \\
    \cmidrule(lr){2-4} \cmidrule(lr){5-8}
    Batch & Fill & Overlap & $T_\mathrm{surface}$ & $E_t^{*}$ & UTS$^{*}$ & $E_f^{*}$ & $\sigma_f^{*}$ \\
    & (mJ/mm) & (\%) & (\si{\celsius}) & (MPa) & (MPa) & (MPa) & (MPa) \\
    \midrule
    S   & 48 & 50 & 168 & $1092$ & $14.4$ & --- & --- \\
    T   & 56 & 50 & 168 & $1675$ & $35.2$ & $1439$ & $47.3$ \\
    U   & 56 & 50 & 172 & $\bm{1678}$ & $\bm{44.4}$ & $\bm{1830}$ & $\bm{60.6}$ \\
    \bottomrule
    \multicolumn{8}{l}{\footnotesize $^{*}$Best-specimen value of tested
      ASTM D638 (\(E_t\), UTS) and D790 (\(E_f\), \(\sigma_f\)) samples.} \\
  \end{tabular}
  \caption{
    Agent optimized print profile settings along with respective tensile and
    flexural properties for the PA12 Blend (25\% PA12 White and 75\% PA12 GF by
    volume) exceeds TDS specified UTS and $\sigma_f$ for PA12 GF displays
    comparable mechanical properties for PA12 White.
  }
  \label{tab:agentic-sls-material-3}
\end{table}

Initial parameter testing of PA12 Blend exhibited visible sintering with an
energy density of 46 mJ/mm. Batch S utilized a fill energy density of 48 mJ/mm,
hotspot overlap of 50\% and surface temperature of 168 \textdegree C printing a
total of 7 ASTM D638 type 4 samples at varying locations within the print bed.
This batch was printed with the excess powder from the parameter testing print
and ended early into the upper stack of ASTM D638 samples 8 - 14 resulting in
only 7 testable samples (Figure \ref{fig:agentic_sls_batch_s_tensile_samples}).
Obtained tensile properties were far below than that of the manufacturer
advertised for either PA12 GF or PA12 White (flexural properties were not
obtained due to print ending early). This however provided a sufficient starting
point for the agentic system to utilize the obtained mechanical properties and
recommend parameter adjustments for the next batch.

\begin{figure}[htbp]
    \centering
    
    \begin{subfigure}[b]{0.34\textwidth}
        \centering
        \includegraphics[width=\textwidth]{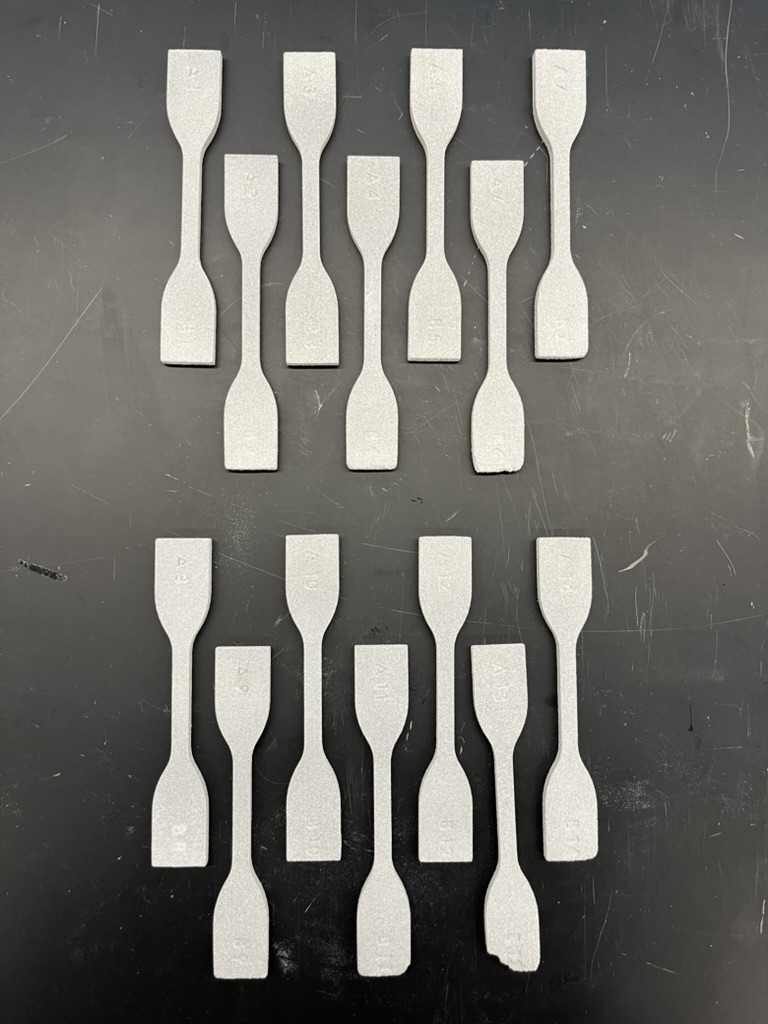}
        \caption{Batch T Tensile}
        \label{fig:agentic_sls_batch_t_tensile_samples}
    \end{subfigure}
    \hfill
    \begin{subfigure}[b]{0.60\textwidth}
        \centering
        \includegraphics[width=\textwidth]{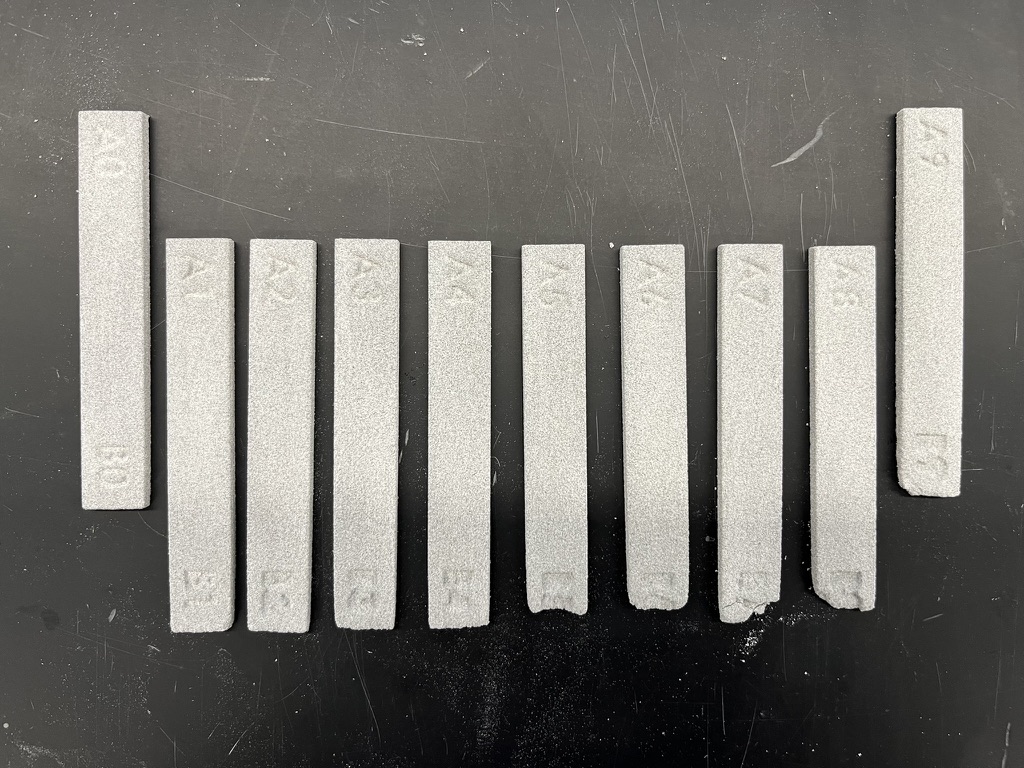}
        \caption{Batch T Flexural}
        \label{fig:agentic_sls_batch_t_flexural_samples}
    \end{subfigure}
    
    \caption{
        Agentically optimized PA12 Blend tensile and flexural samples adjusted
        to print with an elevated fill energy density for Batch T. Portions of
        samples towards the bottom right of the print surface displayed poor
        sintering, as seen with brittle ends in both flexural and tensile
        samples around this area.
    }
    \label{fig:agentic_sls_batch_t_samples}
\end{figure}

For the next print (Batch T) the agentic system was provided the results from
the previous tensile tests and prompted to make adjustments to the print
parameters to improve the resulting mechanical properties. The resulting print
profile utilized a fill energy density of 56 mJ/mm, a hotspot overlap of 50\%,
and a consistent surface temperature of 168 \textdegree C. Samples obtained from
this batch include two stacks of ASTM D638 tensile samples (Figure
\ref{fig:agentic_sls_batch_t_tensile_samples}) and ASTM D790 flexural samples
(Figure \ref{fig:agentic_sls_batch_t_flexural_samples}) arranged in the shown
layout to provide adequate coverage to the print surface. With the adjustment to
fill energy density, tested sample exhibited almost double the previous tensile
properties of Batch S (Table \ref{tab:agentic-sls-material-3}) providing
positive feedback to the agentic system.

\begin{figure}[htbp]
    \centering
    
    \begin{subfigure}[b]{0.34\textwidth}
        \centering
        \includegraphics[width=\textwidth]{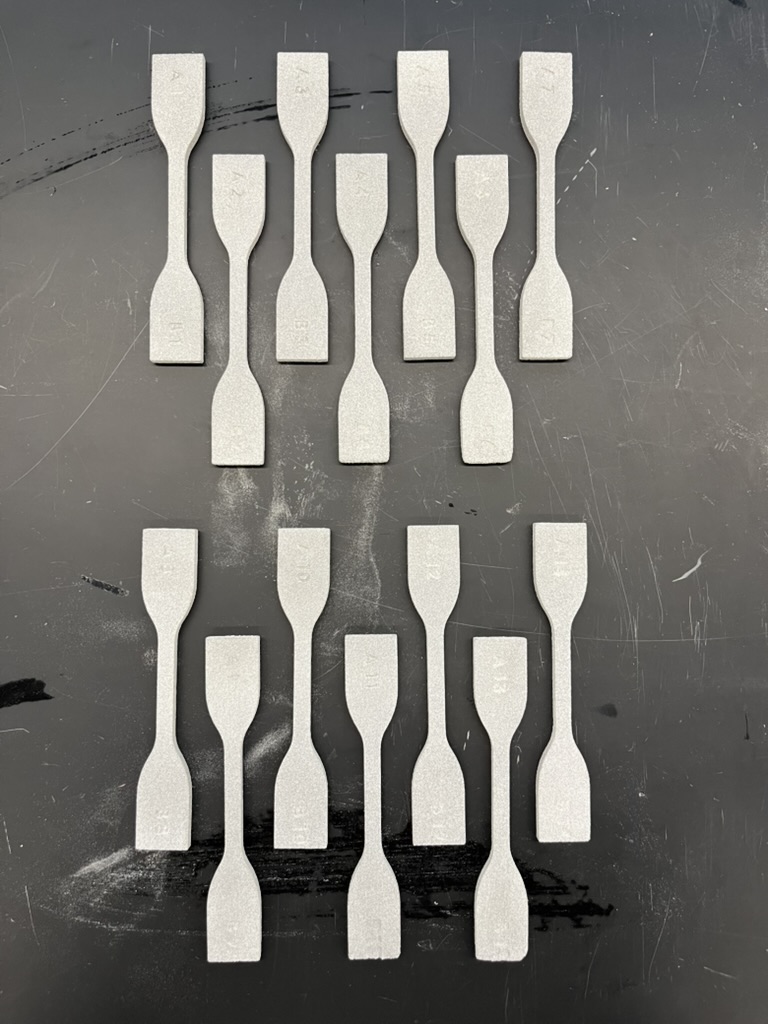}
        \caption{Batch U Tensile}
        \label{fig:agentic_sls_batch_u_tensile_samples}
    \end{subfigure}
    \hfill
    \begin{subfigure}[b]{0.60\textwidth}
        \centering
        \includegraphics[width=\textwidth]{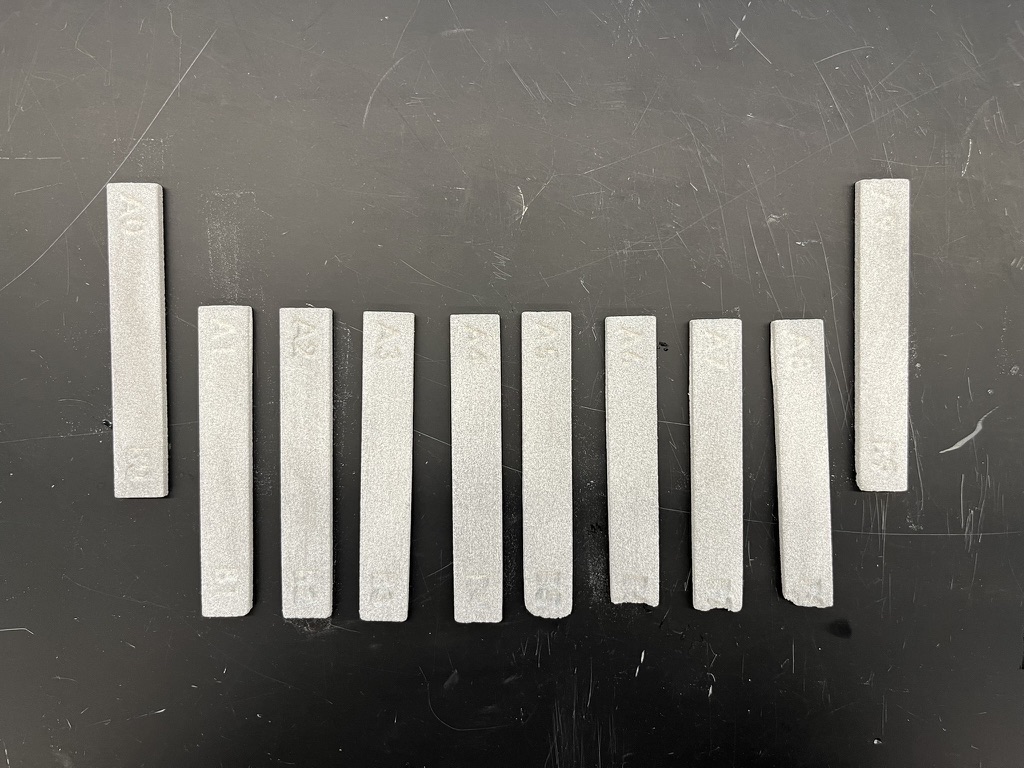}
        \caption{Batch U Flexural}
        \label{fig:agentic_sls_batch_u_flexural_samples}
    \end{subfigure}
    \caption{
        Batch U printed with agentic system recommended surface temperature
        adjustment to 172 \textdegree C resulted in optimal sintering throughout
        all samples with minor brittleness towards the bottom right of the print
        surface.
    }
    \label{fig:agentic_sls_batch_u_samples}
\end{figure}

With these findings, the agentic system was asked to further optimize the print
parameters to provide a more even distribution of ideal tensile and flexural
properties around the print surface (Batch U). For this the previous print
parameters were kept the same with the only adjustment made to the print surface
temperature, elevating it to 172 \textdegree C. Tensile (Figure
\ref{fig:agentic_sls_batch_u_tensile_samples}) and flexural (Figure
\ref{fig:agentic_sls_batch_t_flexural_samples}) exhibited ideal sintering
specifically towards the bottom right of the print surface where the sample
showed brittleness in previous Batch T. Obtained mechanical properties from the
tensile samples showed a slight increase in the tensile modulus of the samples
of around 3 MPa, however the greatest increase was observed in the ultimate
tensile strength of 44.4 MPa from 35.2 MPa, approximately 25\% increase from
that of Batch T. Flexural properties also exhibited similar results increasing
to 1830 MPa for the flexural modulus and 60.6 MPa for the flexural strength.

\begin{figure}[htbp]
    \centering
    
    \begin{subfigure}[b]{0.48\textwidth}
        \centering
        \includegraphics[width=\textwidth]{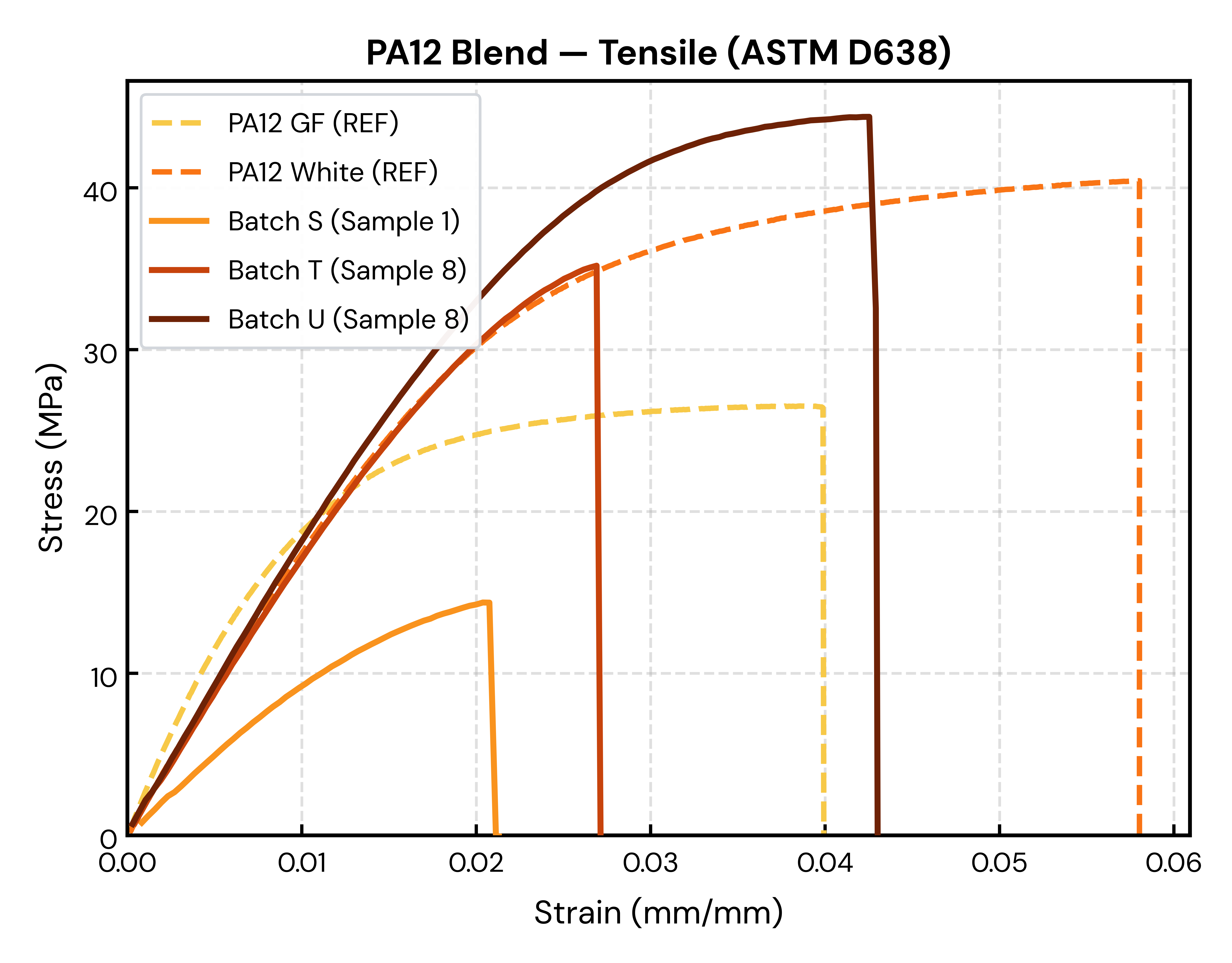}
        \caption{PA12 Blend Tensile Samples}
        \label{fig:agentic_sls_material_3_tensile}
    \end{subfigure}
    \hfill
    \begin{subfigure}[b]{0.48\textwidth}
        \centering
        \includegraphics[width=\textwidth]{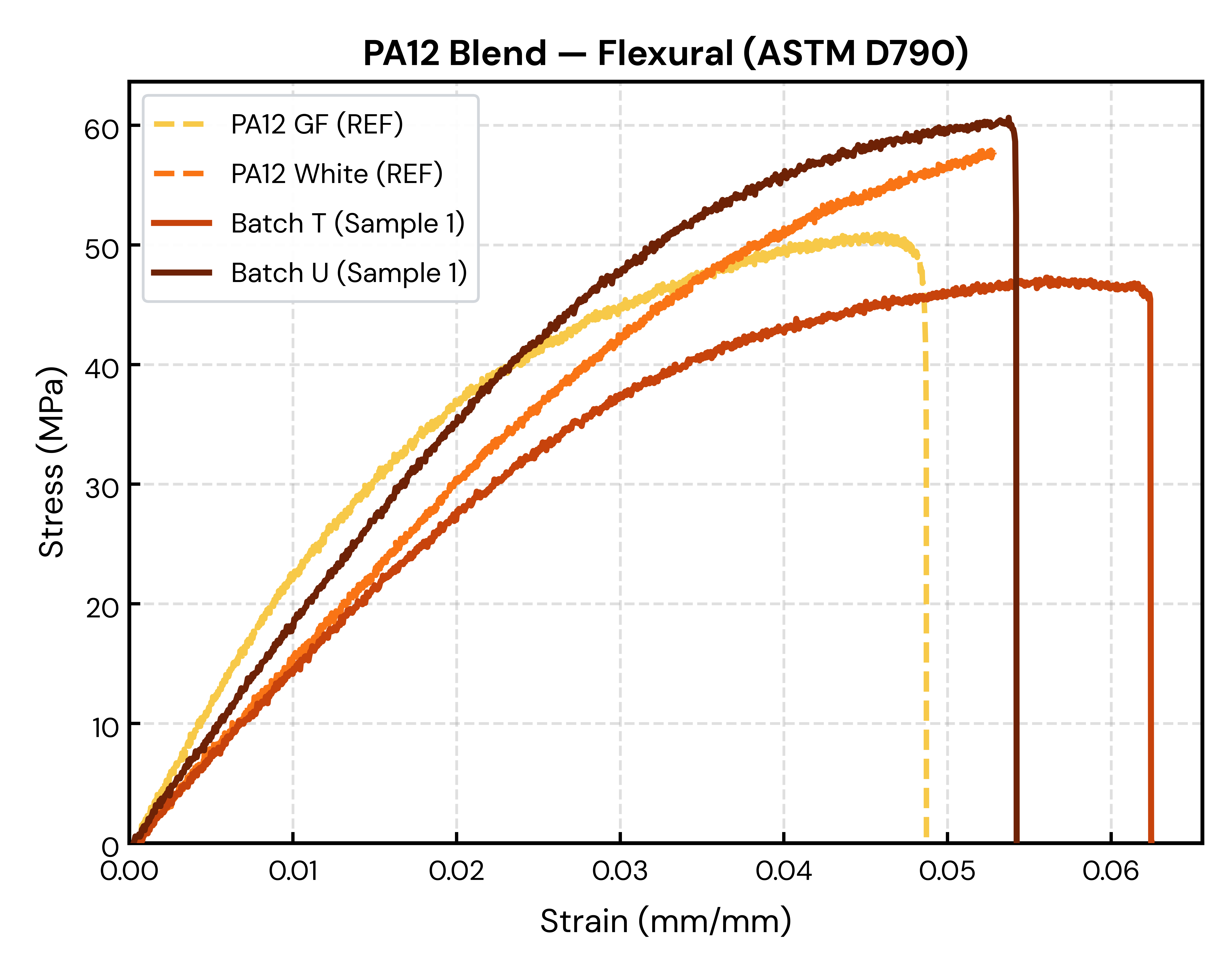}
        \caption{PA12 Blend Flexural Samples}
        \label{fig:agentic_sls_material_3_flexural}
    \end{subfigure}
    \caption{
        PA 12 Blend tensile and flexural properties show gradual improvement
        with each subsequent batch. Best sample from Batch U exhibits greater
        ultimate tensile strength and flexural strength than reference
        materials.
    }
    \label{fig:agentic_sls_material_3_batch_comparisons}
\end{figure}

Reference samples and properties advertised in TDS for Formlabs PA12 White and
PA12 Glass Fiber achieve higher tensile modulus values (TDS 1950 MPa and TDS
2800 MPa respectively) compared to that of PA12 Blend (1678 MPa). However, the
ultimate tensile strength of PA12 Blend (44.4 MPa) is greater than that of PA12
GF (TDS 38.0 MPa and REF 26.5 MPa) but less than that of PA12 White (TDS 47.0
MPa), though greater than that of PA12 White REF at 40.0 MPa (Figure
\ref{fig:agentic_sls_material_3_tensile}). Flexural modulus from PA12 Blend
(1830 MPa) is greater than that of both the REF and TDS of PA12 White (1519 MPa
and 1500 MPa respectively), but less than that of PA12 GF (TDS 2400.0 MPa and
REF 2251.0 MPa). The flexural strength however, is greater than both PA12 White
(TDS 56.0 MPa and REF 58.0 MPa) and PA12 GF (TDS 56.0 MPa and REF 50.9 MPa) at
60.6 MPa for PA12 Blend (Figure \ref{fig:agentic_sls_material_3_flexural}). This
aligns with initial expectations that PA12 Blend would exhibit a mixture of
mechanical properties between PA12 White and PA12 Glass Fiber, sometimes
exceeded reference and TDS properties of both materials in the case of flexural
strength.

\section{Discussion}

\subsection{Baseline Mechanical Properties}

Table \ref{tab:agentic-sls-materials-baseline} provides a listing of the
baseline mechanical properties used for the optimization of materials 1, 2, and
3. Powders manufactured from Formlabs include Reference (REF) samples printed by
Formlabs Formnow service through their Fuse series of printers along with their
respective properties listed in the Technical Data Sheet (TDS). In both PA12
White and PA12 Glass Fiber (GF) it is observed that the TDS exhibits greater
values than that obtained through testing via reference samples. These
deviations are rather minor for Formlabs PA12 White but are significantly
greater for PA12 GF, exhibiting deviations of up to 30\% from that of the
manufacturer's specification. One potential explanation for this could be the
print orientation of the reference sample as visible layer lines on the PA12 GF
D638 sample indicate that the manufacturer printed this sample vertically.

\begin{table}[t]
  \centering
  \begin{tabular}{lcccc}
    \toprule
    & \multicolumn{2}{c}{Tensile} & \multicolumn{2}{c}{Flexural} \\
    \cmidrule(lr){2-3} \cmidrule(lr){4-5}
    Material & $E_t^{*}$ & UTS$^{*}$ & $E_f^{*}$ & $\sigma_f^{*}$ \\
    & (MPa) & (MPa) & (MPa) & (MPa) \\
    \midrule
    Formlabs PA12 White (TDS)           & $\bm{1950.0}$ & $\bm{47.0}$ & $1500.0$ & $56.0$ \\
    Formlabs PA12 White (REF)           & $1920.0$ & $40.0$ & $\bm{1519.0}$ & $\bm{58.0}$ \\
    \midrule
    Formlabs PA12 Glass Fiber (TDS)     & $\bm{2800.0}$ & $\bm{38.0}$ & $\bm{2400.0}$ & $\bm{56.0}$ \\
    Formlabs PA12 Glass Fiber (REF)     & $2577.0$ & $26.5$ & $2251.0$ & $50.9$ \\
    \midrule
    Sinterit PA11 Onyx (TDS)            & $\bm{1680.0}$ & $\bm{55.0}$ & $\bm{1290.0}$ & $\bm{54.2}$ \\
    \bottomrule
    \multicolumn{5}{l}{\footnotesize $^{*}$Best-specimen value of five tested
      per ASTM D638 (\(E_t\), UTS) and D790 (\(E_f\), \(\sigma_f\))} \\
  \end{tabular}
  \caption{
    Baseline tensile and flexural properties obtained from Technical Data Sheet
    (TDS) and manufacturer printed Reference (REF) samples. Reference samples
    printed with Fuse series printers from Formlabs Formnow manufacturing
    service.
  }
  \label{tab:agentic-sls-materials-baseline}
\end{table}

\subsection{Build Layout}
\begin{figure}[htbp]
    \centering
    \includegraphics[width=\textwidth]{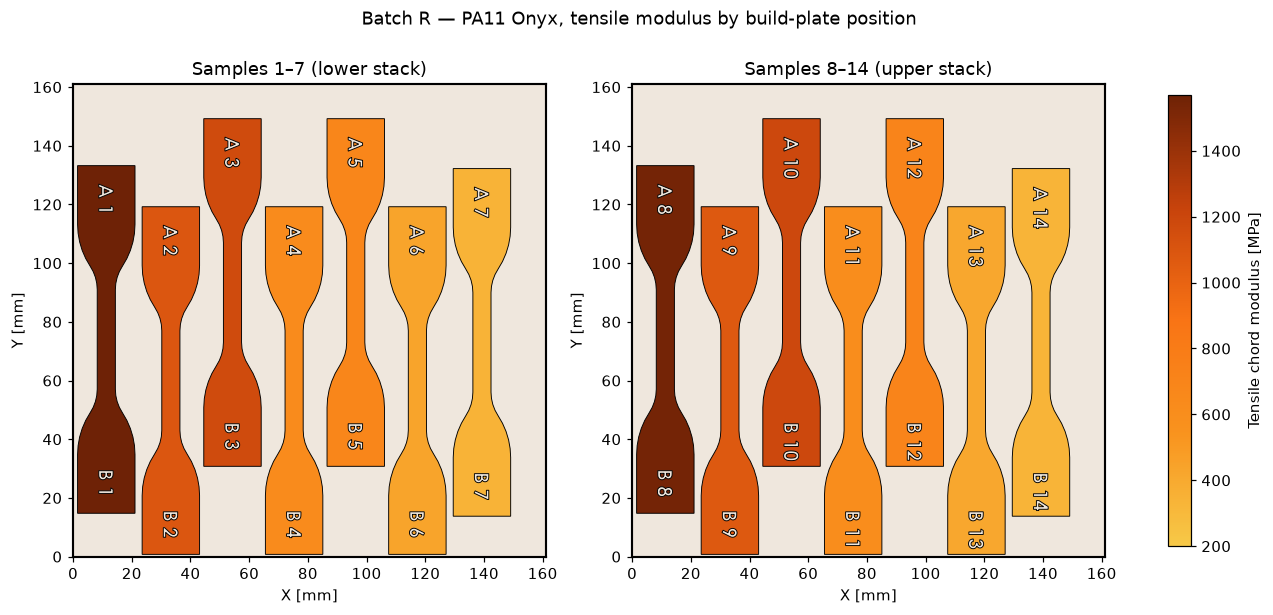}
    \caption{
        Samples 1 and 8 showcase the highest tensile modulus values and are
        closest to the recoater.
    }
    \label{fig:agentic_sls_batch_r_build_layout_tensile_modulus}
\end{figure}

For Material 2 of PA11 Onyx, Batch Q and Batch R achieve tensile modulus values
close to that of the manufacturer published specification, however these
mechanical properties were not observed uniformly across the print bed. This was
reflected in the two stacks of ASTM D638 tensile samples which were printed
where a trend of decreasing mechanical properties were observed from the
recoater side to the overflow side in both stacks. In Batch Q (Figure
\ref{fig:agentic_sls_batch_q_build_layout_tensile_modulus}), the samples with
the highest tensile modulus were found to be closest to the recoater side of the
print chamber where samples closer to the overflow showcased insufficient
sintering, even breaking before testing such is with the case on sample A7 of
Batch Q. Batch R displays the same trend as samples with the highest tensile
modulus are positioned closer to the recoater side of the print chamber (Figure
\ref{fig:agentic_sls_batch_r_build_layout_tensile_modulus}). This may be due to
uneven heating of the print chamber surface where temperatures are higher
towards the recoater and along the Y direction.

\begin{figure}[htbp]
    \centering
    \includegraphics[width=\textwidth]{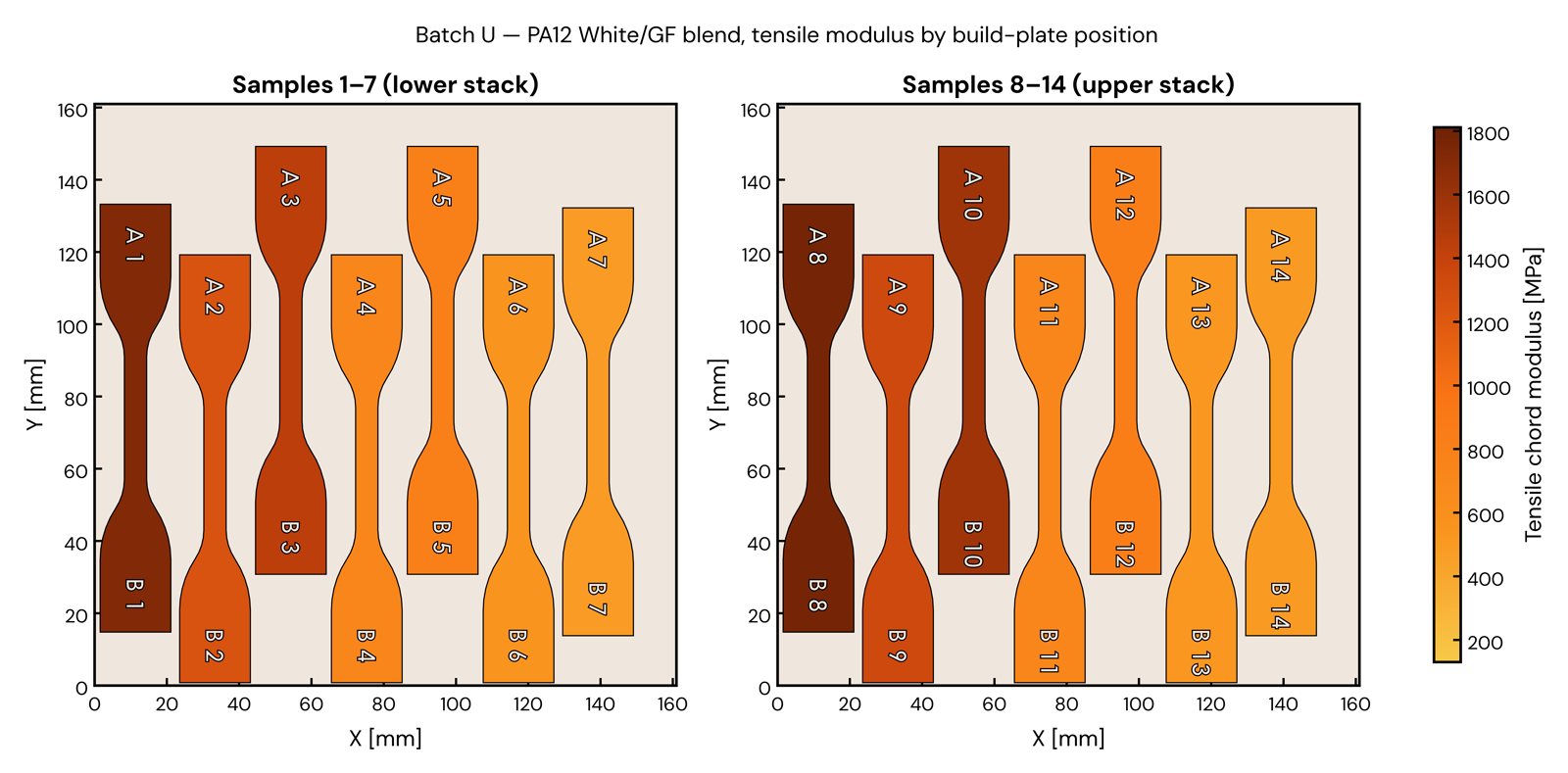}
    \caption{
        Samples 1 and 8 showcase the highest tensile modulus values and are
        closest to the recoater.
    }
    \label{fig:agentic_sls_batch_u_build_layout_tensile_modulus}
\end{figure}

Material 3 confirms this trend with Batch S, T, and U for Nylon 12 Blend. Batch
S (Figure \ref{fig:agentic_sls_batch_s_build_layout_tensile_modulus}) and Batch
T (Figure \ref{fig:agentic_sls_batch_t_build_layout_tensile_modulus}) show a
similar trend in mechanical properties with placement along the print surface.
Batch U produced the best results and does show a more even distribution of
tensile modulus however a gradient of properties is still visible (Figuke
\ref{fig:agentic_sls_batch_u_build_layout_tensile_modulus}).

\section{Conclusion}

With initial calibration on Material 1 (PA12 GF) and further tuning with
Material 2 (PA11 Onyx), the agentic system is able to tune selective laser
sintering process parameters, optimizing for tensile and flexural properties,
achieving mechanical properties that exceed that of the reference samples on
Material 3 (PA12 Blend). Using knowledge from previous builds and their
respective mechanical properties and minimal guidance from the user, the agentic
system was able to optimize process parameters over a small number of iterations
to achieve comparable TDS specified mechanical properties for various materials.
This work showcases the ability for an agentic system to continually learn from
updated data, enabling the intelligent automation of complex tasks such as
process parameter optimization for selective laser sintering.

\clearpage
\appendix

% Redefine \thesection to include "Appendix"
\renewcommand{\thesection}{Appendix \Alph{section}}

\section{Manufactured Samples}
\begin{figure}[htbp]
    \centering
    \includegraphics[width=0.5\textwidth]{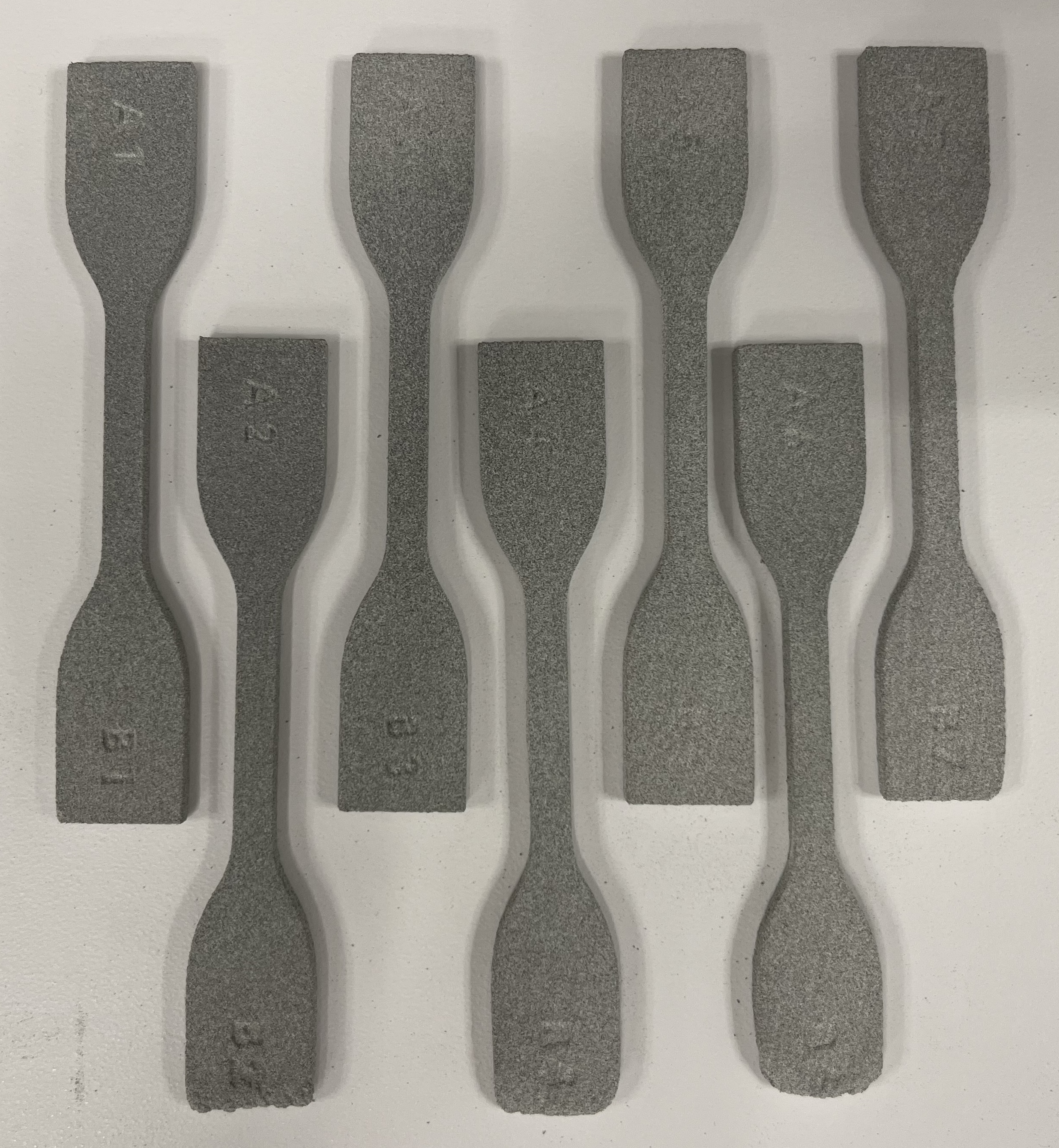}
    \caption{
        Initial set of tensile samples (Batch S) printed with Nylon 12 Blend
        (25\%  PA12 GF and 75\% PA12 White by volume) exhibits a relatively
        lighter color than Material 1 or Material 2.
    }
    \label{fig:agentic_sls_batch_s_tensile_samples}
\end{figure}

\section{Layout Dependent Tensile Properties}
\begin{figure}[htbp]
    \centering
    \includegraphics[width=\textwidth]{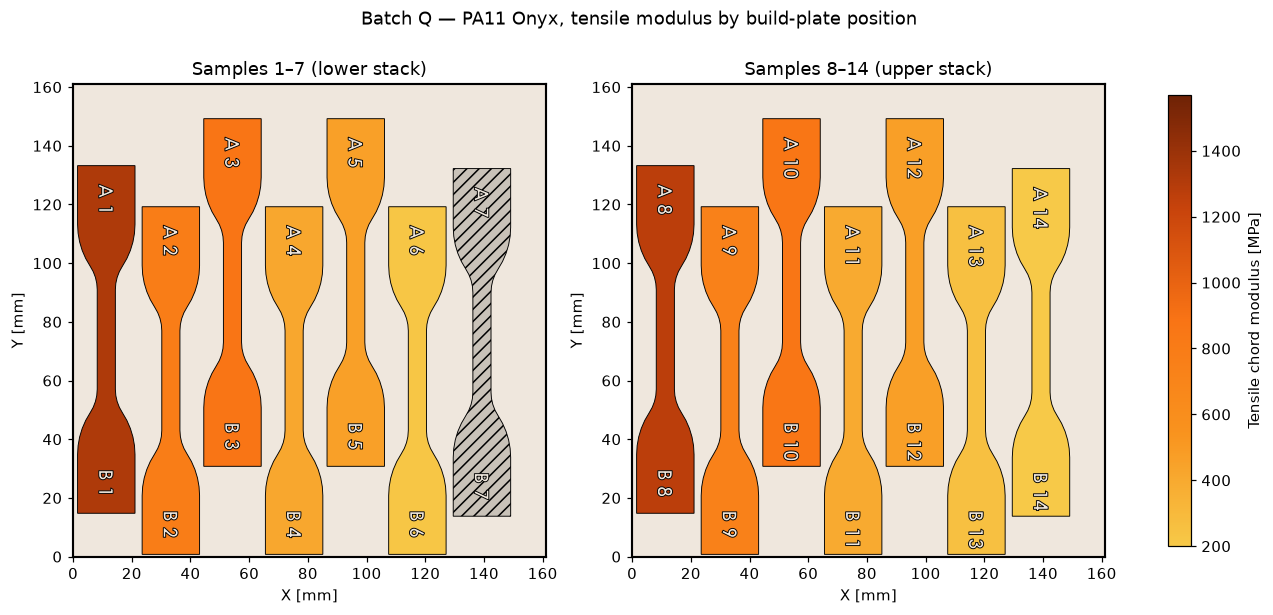}
    \caption{
        Batch Q Samples printed with PA11 Onyx showcased a trend of weaker
        mechanical properties with increasing x and decreasing y positioning on
        the print chamber. Sample 7 displayed low stiffness and fractured before
        tensile load values recorded.
    }
    \label{fig:agentic_sls_batch_q_build_layout_tensile_modulus}
\end{figure}

\begin{figure}[htbp]
    \centering
    \includegraphics[width=\textwidth]{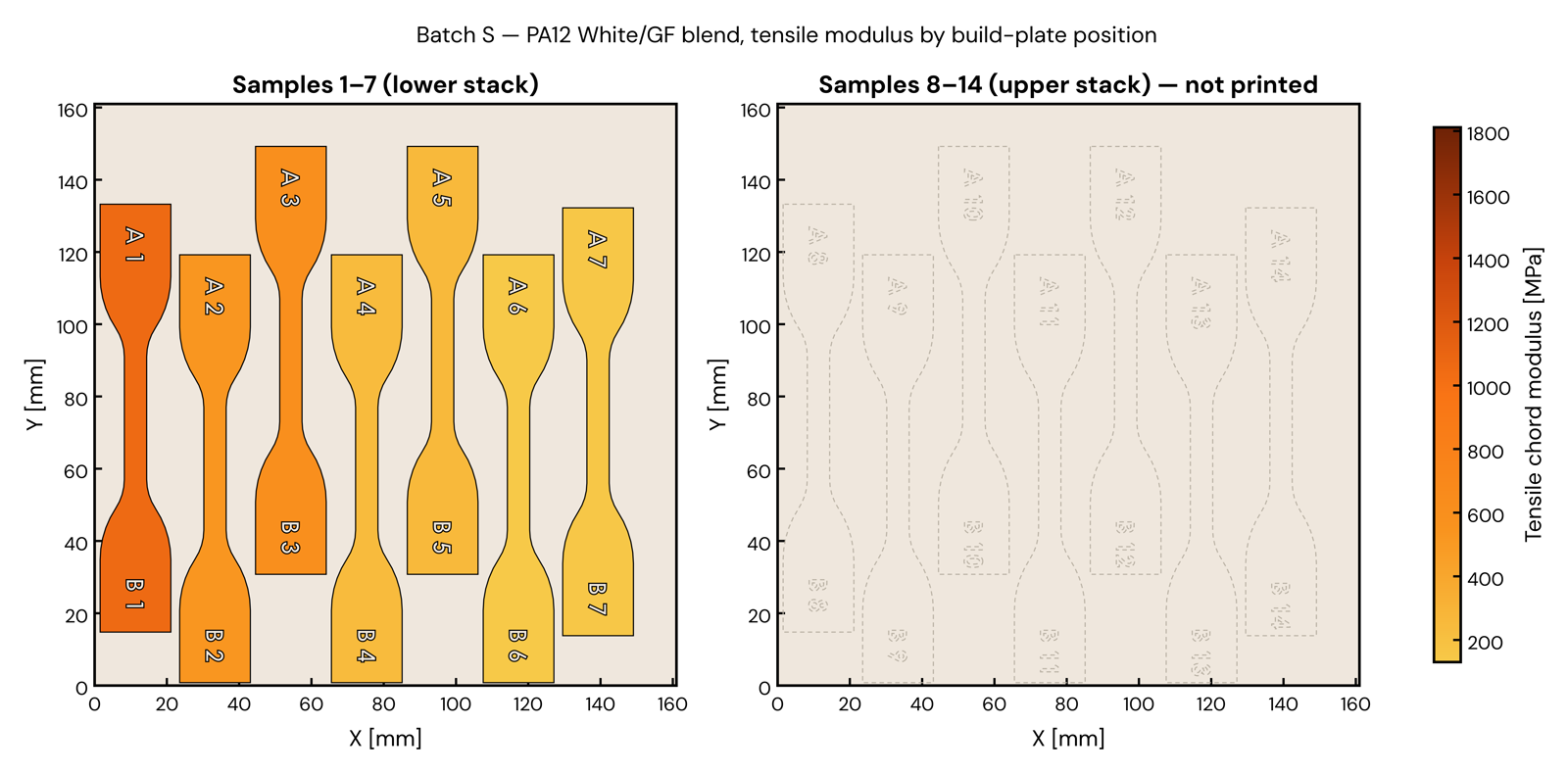}
    \caption{
        Batch S Samples printed with PA12 Blend displays similar trend of
        varying mechanical properties with respect to placement along the print
        surface. Lower stack was only printed due to run out of powder during
        print.
    }
    \label{fig:agentic_sls_batch_s_build_layout_tensile_modulus}
\end{figure}

\begin{figure}[htbp]
    \centering
    \includegraphics[width=\textwidth]{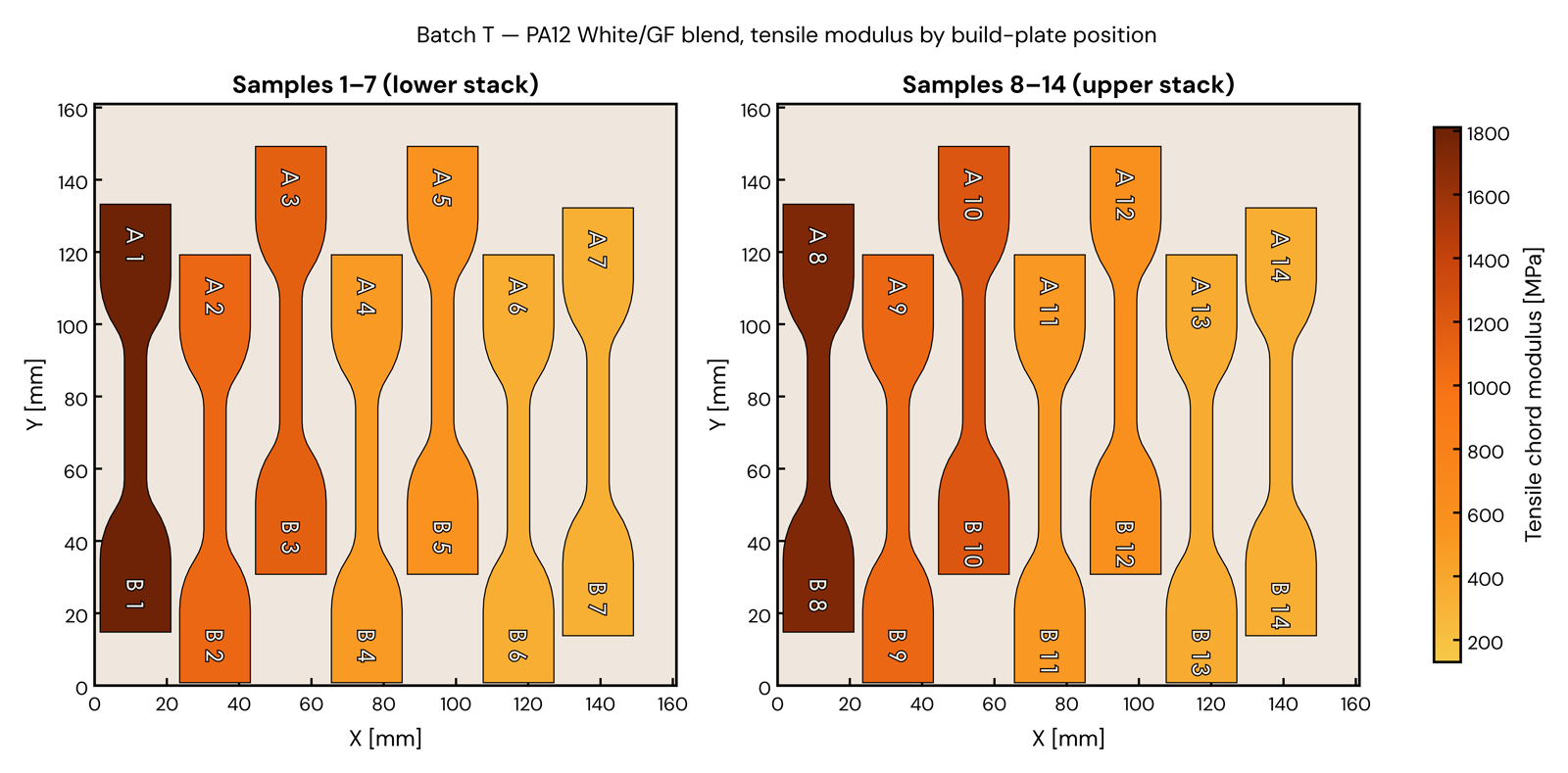}
    \caption{
        Batch T Samples printed with PA12 Blend exhibit greater tensile
        properties with the similar dependence on location within the print
        surface.
    }
    \label{fig:agentic_sls_batch_t_build_layout_tensile_modulus}
\end{figure}

% \bibliography{achemso-demo}
\bibliography{references}

\end{document}